\documentclass[12pt]{article}

\usepackage{placeins}

\usepackage{newtxtext,newtxmath}

\usepackage{graphicx}

\usepackage[letterpaper,margin=1in]{geometry}

\renewenvironment{abstract}
	{\quotation}
	{\endquotation}

\date{}

\makeatletter
\renewcommand{\fnum@figure}{\textbf{Figure \thefigure}}
\renewcommand{\fnum@table}{\textbf{Table \thetable}}
\makeatother

\usepackage{scicite}

\usepackage{url}

\def\scititle{
	The cost of being seen as AI in group bargaining
}
\title{\bfseries \boldmath \scititle\\[0.5em]
{\normalfont\normalsize
}}

\author{
	Tahmineh~Godazgar$^{1}$,
	Nico~Mutzner$^{2}$,
	Taha~Yasseri$^{1,3\ast}$\and
	\small$^{1}$Centre for Sociology of Humans and Machines, Trinity College Dublin \\ \small and Technological University Dublin, Dublin, Ireland\and
	\small$^{2}$Department of Sociology, University of Zurich, Zurich, Switzerland\and
	\small$^{3}$School of Mathematics and Statistics, University College Dublin, Dublin, Ireland\and
	\small$^\ast$Corresponding author. Email: taha.yasseri@tcd.ie
}

\begin{document}

\maketitle

\begin{abstract} \bfseries \boldmath
As artificial intelligence (AI) becomes embedded in collective decision-making, people's perceptions of AI may shape who reaches agreement and what they receive. We studied this in a repeated three-player majoritarian bargaining game with 2,234 participants. Players saw each opponent identified as Human or AI. The instructions described AI as acting either independently or on behalf of a human. Human participants made every decision. In mixed groups, players presented as AI received fewer points in opening offers, were included in fewer final agreements and earned about 18\% less per round than those presented as Human. Smaller payoffs in agreements that included them accounted for most of the observed gap. Further analysis found a payoff gap even when all three players agreed. Social perceptions of AI may therefore shape who benefits from collective decisions, with potential costs for people represented by AI.
\end{abstract}



\section*{Introduction}

\noindent
When groups bargain over a shared resource, they decide who joins an agreement and how to divide the gains. Proposers often secure a larger share, and a deal can leave some participants with nothing \cite{camerer2003,baron1989}. As AI systems take on roles in these decisions, both as agents and as delegates, the way others perceive them may affect who gets included and what they receive. Whether a counterpart is taken to be human or AI can shape expectations about their behaviour, agency and standing in the group \cite{seaver2017,liu2024}.

In experimental games, people tend to be less prosocial towards AI counterparts \cite{zhou2026}. Proposed explanations include greater self-interest, a willingness to exploit cooperative AI, and weaker reciprocity \cite{march2021,karpus2021,upadhyaya2023}. In some settings, describing the same behaviour as AI-generated reduces cooperation or trust and changes judgements of fairness \cite{ishowo2019,dvorak2025}. Fair behaviour, shared cooperative norms, and the information available about individual partners can lessen these differences \cite{wang2026,mutzner2026,makovi2023}. People also judge similar offers differently depending on who made them and attribute different mental states to AI and human counterparts \cite{demelo2014,niszczota2026,lee2021,swiderska2019,sanfey2003}. These studies point to several reasons why being seen as AI or represented by AI could matter in bargaining.

Group bargaining also allows players to choose among potential partners. A third player makes it possible to distinguish two outcomes: who is included in an agreement and what they receive. Any two of the three players can reach an agreement, so a player seen as AI may be left out or may join the agreement on less favourable terms \cite{baron1989,frechette2003}. In experiments with the same bargaining structure, Baranski, Geraldes and colleagues found that women earned less and most of the gap reflected exclusion from coalitions \cite{baranski2024}. Research on partner choice suggests that disclosing an AI identity can affect inclusion, with selection differences changing as players learn from repeated interaction \cite{jiang2025}. Work on fairness towards AI counterparts also gives reason to expect differences in the terms offered \cite{demelo2014,fu2026}. Prior group studies have examined AI in assigned roles, as advisers or delegates, or as possible partners in tasks focused on partner choice \cite{fernandez2022,shirado2017,zhu2026,jiang2025}. Here we examine a setting where players choose whom to reach agreement with and how to split a common surplus.

It may also matter whether AI is seen to act independently or on someone's behalf. When people understand that a human benefits from an AI agent's earnings, they may respond more favourably to it in bargaining \cite{ozkes2024,fu2026}. Describing a delegate in intentional terms can also bring responses closer to those towards humans \cite{globig2026}. Other work suggests that a penalty can extend to the person represented by the AI. Fairness, trust and cooperation can decline when a model makes a partner's decisions, and people may judge its offers as less appropriate even when it acts on someone's behalf \cite{dvorak2025,niszczota2026}. People's perceptions of an AI can also shape how they treat it \cite{treiman2026}. We therefore compare two framings, one describing an autonomous AI agent and the other an AI delegate for a human. In what follows, \emph{Human} and \emph{AI} refer to how we presented the players to their opponents; all players were human participants. We preregistered five hypotheses (\url{https://osf.io/w6xfd}):

\textit{H1 (Label disadvantage):} We expected AI players to earn lower total payoffs than Human players.

\textit{H2 (Partner preference):} When both labels were available among the opponents, we expected players to form coalitions more often with Human partners than with AI partners.

\textit{H3 (First-offer bias):} We expected players to allocate more points to Human opponents than to AI opponents in their own first proposal in a round.

\textit{H4 (Exploitation):} We expected players to retain more points in coalitions with AI partners than in coalitions with Human partners.

\textit{H5 (Representation advantage):} We expected the disadvantage across outcomes to be smaller when AI players were described as delegates of humans than when they were described as autonomous AI agents.

\begin{figure}[!htbp]
	\centering
	\includegraphics[width=1.0\textwidth]{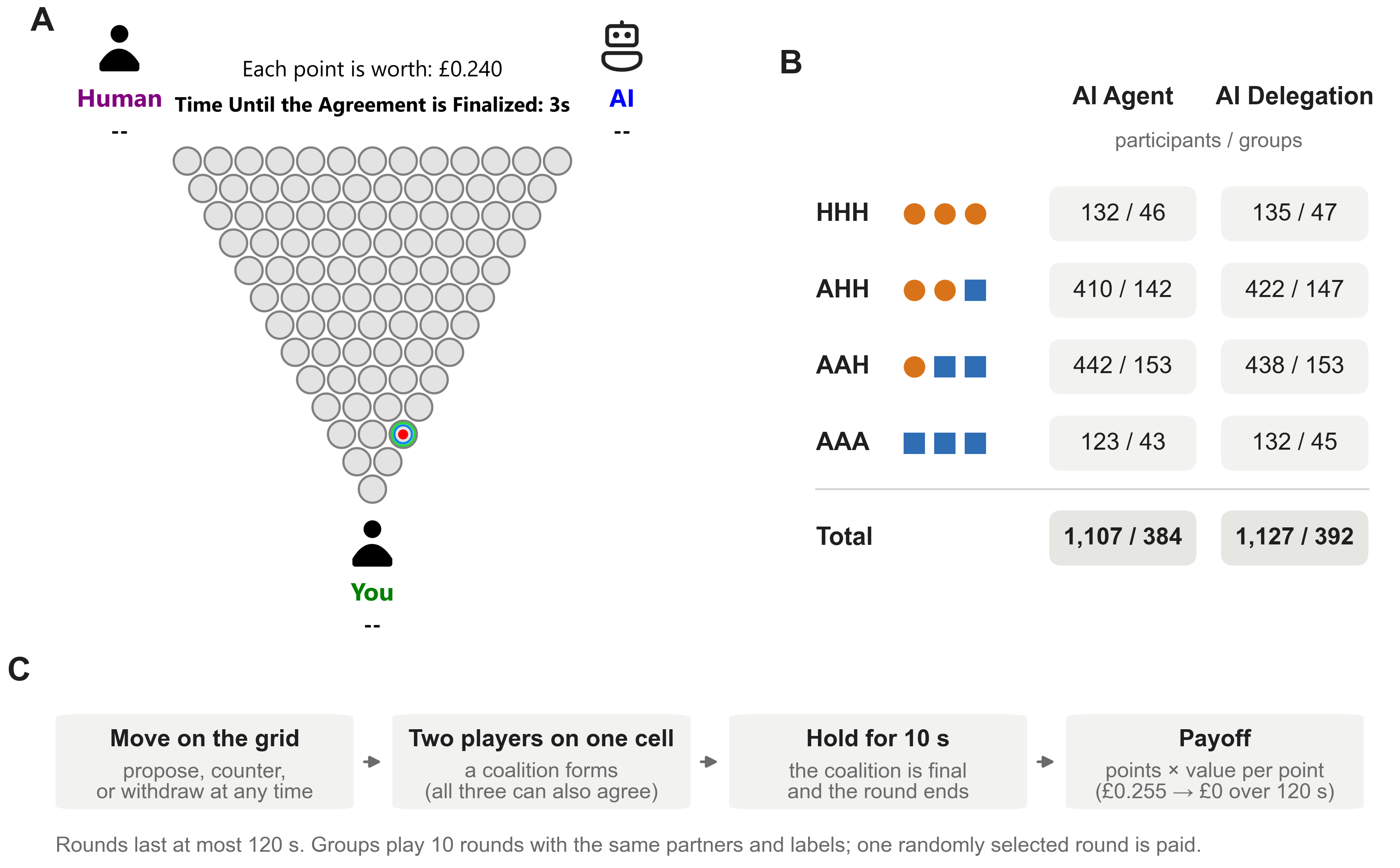}
	\caption{\textbf{Experimental design.} (\textbf{A}) Bargaining interface. Each grid cell divides 12 points among the three players. The screen identifies each opponent as Human or AI and marks the player's own position as ``You''. All players were human participants; the labels describe how they appeared to opponents. (\textbf{B}) Retained participants and groups in each of the four label compositions crossed with the two AI framings. (\textbf{C}) Structure of a round. Labels, framing and group membership remained fixed over ten rounds. One randomly selected round determined each participant's bonus.}
	\label{fig:design}
\end{figure}

We tested these hypotheses in a preregistered online bargaining experiment, with an analytical sample of 2,234 participants drawn from 776 groups (Fig.~\ref{fig:design}, Tables~\ref{tab:attrition} and \ref{tab:sample}). Each group was formed as a triad, although exclusions left some groups with fewer than three analysable participants. Each group kept the same members, labels and AI framing for ten rounds. The four group compositions were HHH, AHH, AAH and AAA, with H and A denoting players presented as Human and AI, respectively.

In each round, at least two players had to agree on a split of 12 points and hold it for ten seconds to finalise the agreement. This allowed us to observe both two-player and three-player agreements, and to compare what players received in each. Participants had a monetary incentive to reach a favourable split, with a bonus based on one randomly selected round. The recorded moves identify each player's first proposal and subsequent moves. Because all decisions were made by humans using the same rules, the design isolates responses to how the players were presented.

\section*{Results}
In mixed groups, players presented as AI earned about 18\% less per round than those presented as Human. They received fewer points in opening offers, were included in fewer final agreements, and earned less when included. The payoff gap was also present in agreements accepted by all three players. It appeared when AI players were described as autonomous agents and when they were described as acting on behalf of humans.

H1--H5 address the preregistered predictions. Additional analyses of group and opponent composition and of three-player agreements were not preregistered and are treated as exploratory. These analyses describe where the payoff gap appears beyond the main tests. Detailed comparisons and additional results are reported in Tables~\ref{tab:h1means}--\ref{tab:demorobust} and Figs.~\ref{fig:slabels}--\ref{fig:ssurvey}.

\subsection*{AI players earn less in mixed groups}

\begin{figure}[!htbp]
	\centering
	\includegraphics[width=1.0\textwidth]{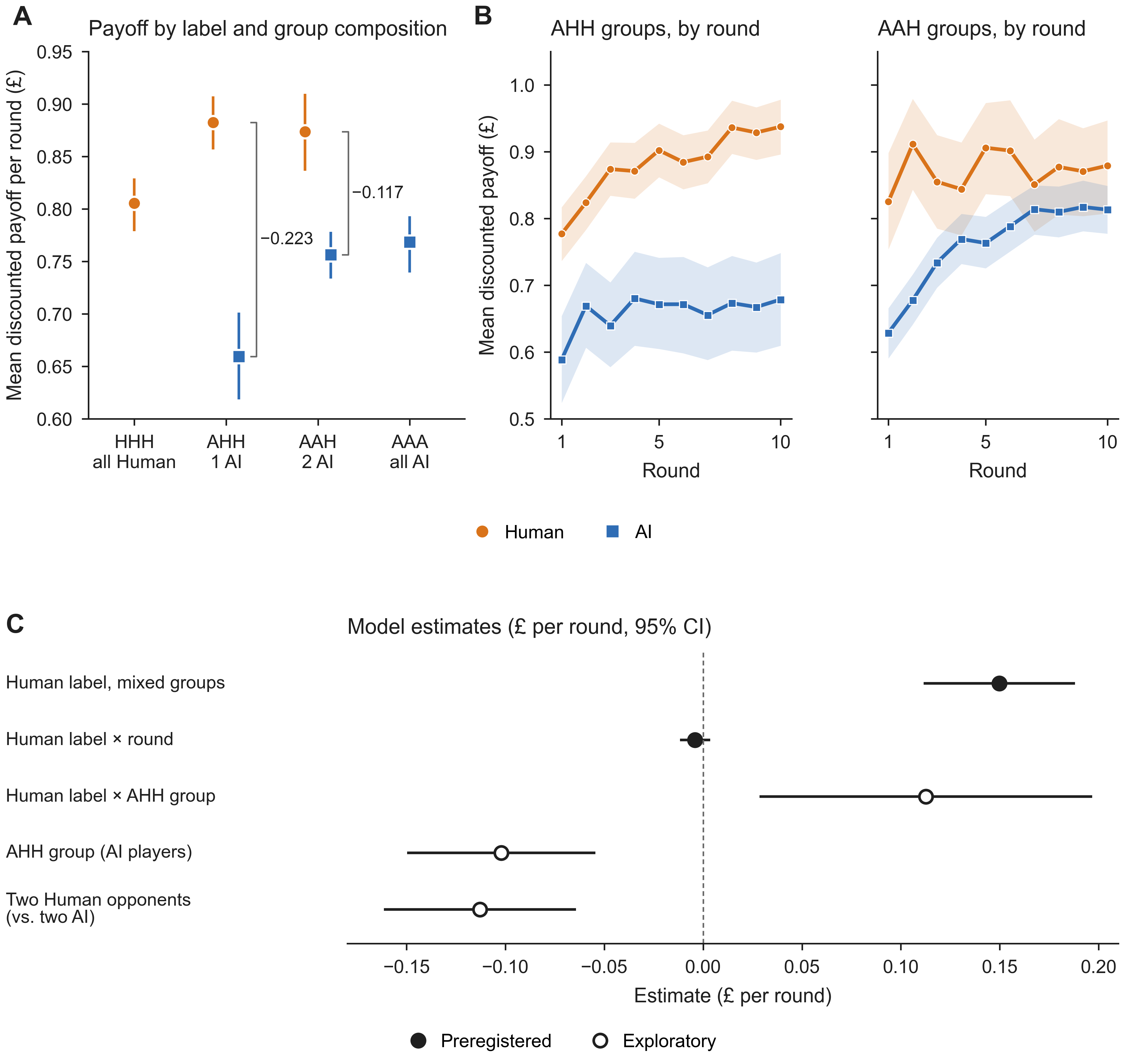}
	\caption{\textbf{The payoff gap was larger when most players were presented as Human.} (\textbf{A}) Participant-level mean discounted payoff per round by player label in each group composition, with rounds in which the player was not part of the agreement counted as zero. (\textbf{B}) Mean discounted payoff by round in AHH and AAH groups. (\textbf{C}) Mixed-effects estimates in pounds per round, with 95\% confidence intervals. Filled markers show the preregistered Hypothesis 1 comparisons. Open markers show exploratory estimates for group and opponent composition (Table~\ref{tab:context}).}
	\label{fig:penalty}
\end{figure}

In mixed groups, the mean discounted payoff per round was \pounds0.725 for AI players and \pounds0.879 for Human players (Fig.~\ref{fig:penalty}A and Table~\ref{tab:h1means}). The raw difference between these means was \pounds0.154 per round (Cohen's $d = 0.44$). A mixed-effects model with standard errors clustered by group estimated a similar Human advantage of \pounds0.150 per round (95\% confidence interval (CI) [0.112, 0.188], $p < 0.001$), supporting H1. The observed and model-estimated gaps therefore point to the same disadvantage.

The size of the gap varied with group composition. It was nearly twice as large with a Human majority (\pounds0.223 per round) as with an AI majority (\pounds0.117). A model comparing these compositions estimated a \pounds0.113 larger gap with a Human majority (95\% CI [0.029, 0.197], $p = 0.009$, Fig.~\ref{fig:penalty}C and Table~\ref{tab:context}). Mean payoffs were \pounds0.805 in all-Human groups and \pounds0.768 in all-AI groups (Fig.~\ref{fig:penalty}A). In that between-group comparison, all three labels change together.

To compare the focal player's label while holding both opponents' labels fixed, we ran a second model comparing Human and AI players facing the same opponent-label combinations (Table~\ref{tab:context}). The estimated Human advantage was \pounds0.147 per round against two Human opponents (95\% CI [0.097, 0.196]), \pounds0.121 against one opponent of each label ([0.087, 0.155]), and \pounds0.100 against two AI opponents ([0.056, 0.144]). Thus, the point estimate favoured Human players in every opponent configuration.

The gap was also evident in undiscounted points. The pooled label-by-round interaction provided little evidence of a systematic change in the gap over the ten rounds ($b = -0.004$, 95\% CI [$-0.012$, 0.004], $p = 0.295$, Fig.~\ref{fig:penalty}B, C).

The overall gap reflects differences in inclusion and in agreement terms. In mixed groups, AI players were part of the final agreement in 83.5\% of rounds, compared with 88.2\% for Human players. In rounds in which they were included, they received \pounds0.873 and \pounds0.996 on average, respectively. Weighting each participant-round equally, a descriptive decomposition attributed 71\% of the resulting \pounds0.150 gap (\pounds0.154 when each participant is weighted equally) to differences in payoffs among included players and 29\% to differences in inclusion rates (Table~\ref{tab:exclusion}). The payoff contrast among included players is descriptive because inclusion is itself a bargaining outcome.

\subsection*{The payoff gap persists in three-player agreements}

\begin{figure}[!htbp]
	\centering
	\includegraphics[width=1.0\textwidth]{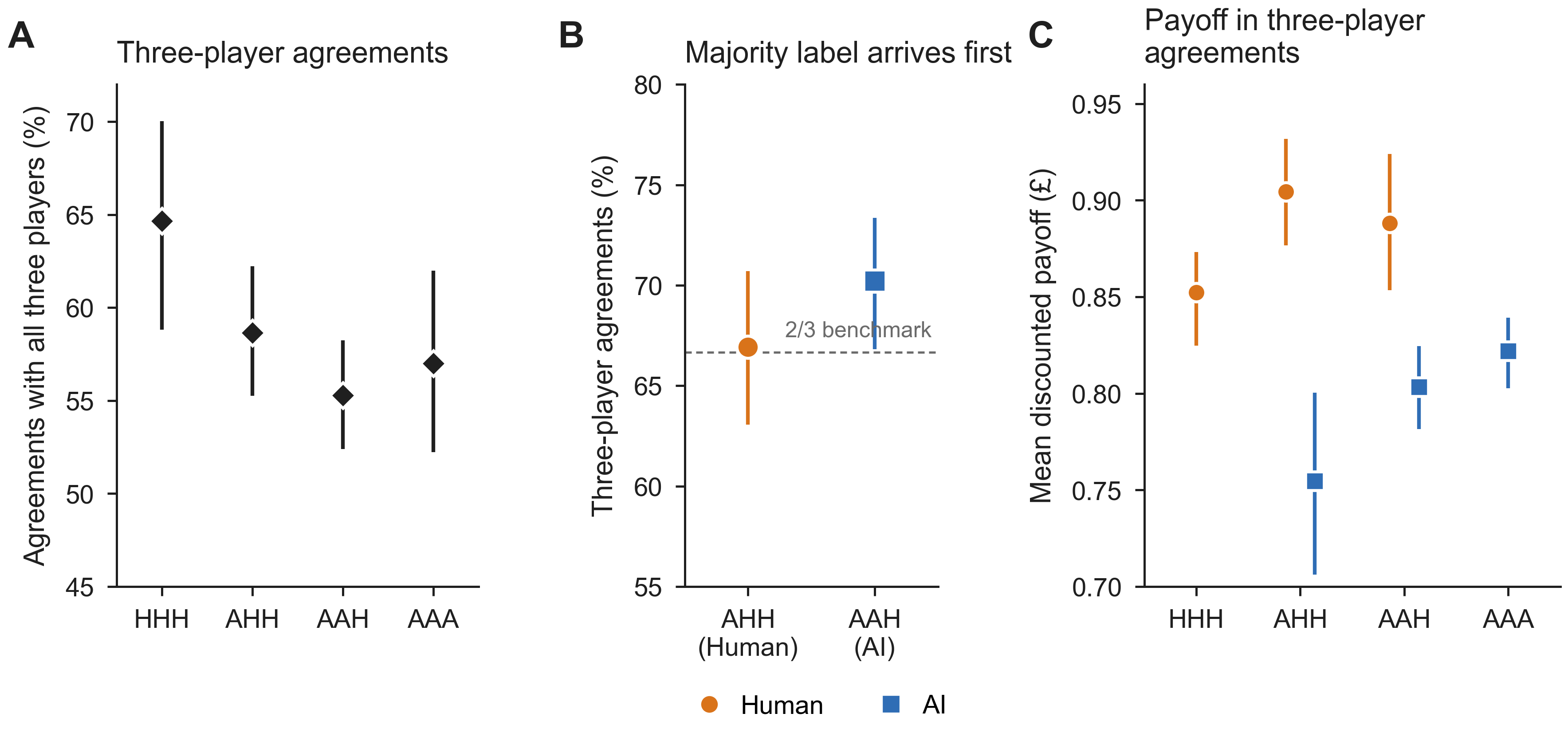}
	\caption{\textbf{Players presented as AI earned less in three-player agreements.} (\textbf{A}) Percentage of coalition rounds ending in a three-player agreement, by group composition. (\textbf{B}) Percentage of three-player agreements in mixed groups in which the first player to arrive at the final cell carried the majority label. The dashed line marks the two-thirds share expected if arrival order were unrelated to label. (\textbf{C}) Mean discounted payoff (\pounds) within three-player agreements by player label and group composition. Error bars are 95\% confidence intervals. The three-player-agreement analysis was not preregistered.}
	\label{fig:grand}
\end{figure}

In mixed groups, AI players earned less even when all three players joined the final agreement. The difference was present under both Human and AI majorities (Fig.~\ref{fig:grand}C and Table~\ref{tab:grandpay}). After accounting for group composition and round, the estimated Human advantage was \pounds0.117 per round (95\% CI [0.073, 0.161], $p < 0.001$). Agreement by every player therefore coexisted with unequal payoffs across the two labels.

Across all four group compositions, agreements involving all three players accounted for 57.8\% of the 6,638 coalition rounds with complete move histories (Fig.~\ref{fig:grand}A and Table~\ref{tab:grandprev}).

\subsection*{Human players more often agreed with one another}

\begin{figure}[!htbp]
	\centering
	\includegraphics[width=1.0\textwidth]{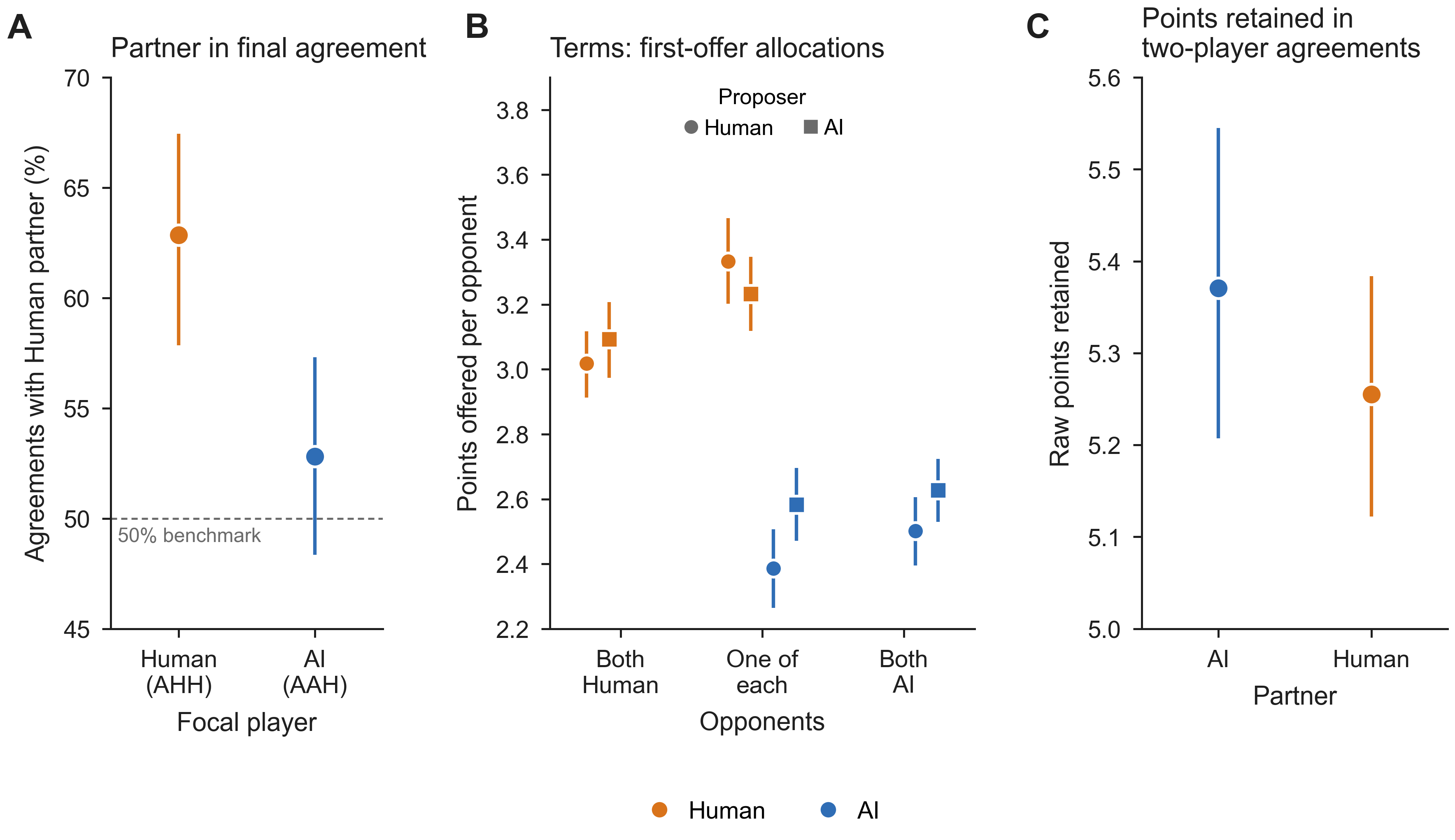}
	\caption{\textbf{Partner labels in final agreements, first offers and retained points.} (\textbf{A}) Percentage of finalised two-player agreements that paired a focal player with the Human opponent, among focal players facing one opponent of each label (Hypothesis 2, Table~\ref{tab:sh1}). (\textbf{B}) Points offered to each opponent in a player's first proposal of a round, averaged within participants and shown by the two opponents' labels (Hypothesis 3, Table~\ref{tab:sh2}). (\textbf{C}) Mean raw points retained in finalised two-player agreements, by partner label (Hypothesis 4, Table~\ref{tab:sh3}). Error bars show 95\% confidence intervals.}
	\label{fig:routes}
\end{figure}

When Human players faced one opponent of each label, 62.9\% of the finalised two-player agreements they joined were with the Human opponent (1,115 of 1,774, Fig.~\ref{fig:routes}A). For AI players facing the same mix of opponents, the corresponding share was 52.8\% (909 of 1,721). A logistic model fitted with generalised estimating equations (GEE), with one observation per participant-round and clustered by bargaining group, estimated an above-50\% probability for Human players at the midpoint of the ten rounds. For AI players, the estimate was compatible with 50\% (Table~\ref{tab:sh1}). The estimated contrast between the two focal-player labels was $b = 0.424$ (95\% CI [0.151, 0.697], $p = 0.002$).

This pattern partly supports H2. The difference concerns which partners ultimately reached a deal together. These agreements were reached jointly by players. They also compare two group compositions, because a Human focal player faces one opponent of each label in AHH groups, and an AI focal player does so in AAH groups.

In exploratory pooled comparisons, initial coalitions broke in 13.4\% of Human--Human pairings and 18.0\% of AI--AI pairings (Fig.~\ref{fig:sbroken}). These descriptive comparisons also vary in group composition.

\subsection*{AI players receive smaller opening offers}

When players faced one opponent of each label, they offered the AI opponent fewer points in their own first proposal of the round. Human proposers offered 3.333 points on average to Human opponents and 2.386 to AI opponents. AI proposers offered 3.233 and 2.583 points, respectively (Fig.~\ref{fig:routes}B). Across all proposers in mixed groups, including those facing two opponents with the same label, the mixed-effects model estimated a 0.728-point premium for Human opponents (95\% CI [0.609, 0.847], $p < 0.001$, Table~\ref{tab:sh2}), supporting H3. The label difference was therefore visible in what players proposed and in the agreements that ultimately formed.

In the descriptive comparisons, the offer gap averaged about 0.80 points when the two opponents carried different labels and 0.49 points when they carried the same label (Fig.~\ref{fig:routes}B). The latter estimate compares offers across groups with two Human opponents and those with two AI opponents. The pooled offer gap showed little evidence of a systematic change over the ten rounds ($p = 0.380$, Table~\ref{tab:sh2}).

\subsection*{Retained points did not clearly differ by partner label}

Among focal players who faced one opponent of each label and reached a two-player agreement, those with an AI partner retained 5.370 raw points on average, compared with 5.255 for those with a Human partner (Fig.~\ref{fig:routes}C). In the preregistered pooled specification (3,495 participant-round observations), having a Human partner was associated with 0.179 fewer points (95\% CI [$-0.386$, 0.027], $p = 0.089$). The corresponding estimate for discounted payoff was \pounds0.035 lower (95\% CI [$-0.083$, 0.013], $p = 0.152$, Table~\ref{tab:sh3}). Both estimates point in the direction predicted by H4, but the intervals include zero, so the results do not support H4 at the two-sided 5\% level. Under the preregistered one-sided test, the estimate for retained points would be significant ($p = 0.044$), while the estimate for discounted payoff would not ($p = 0.076$). The estimates describe the deals that formed; they do not show what a player would have earned by choosing a different partner.

\subsection*{The payoff gap appeared under both AI framings}

\begin{figure}[!htbp]
	\centering
	\includegraphics[width=1.0\textwidth]{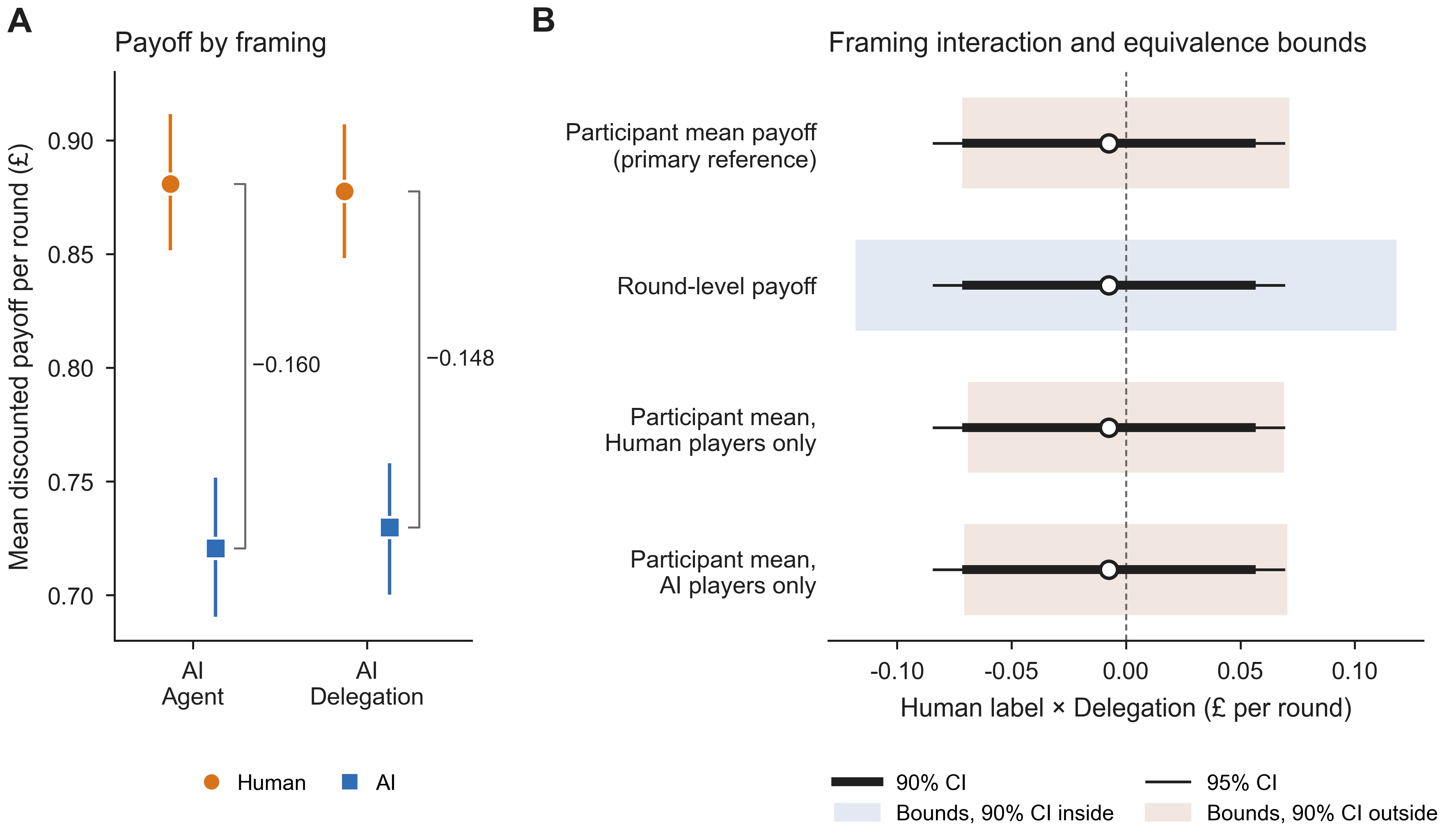}
	\caption{\textbf{Payoffs by label under the AI Agent and AI Delegation framings.} (\textbf{A}) Participant-level mean discounted payoff per round in mixed groups. Rounds outside the final agreement count as zero. (\textbf{B}) Estimated interaction between the Human label and Delegation framing. Thick and thin lines show 90\% and 95\% confidence intervals, respectively. The shaded regions show the equivalence bounds corresponding to 0.20 standard deviations for four reference standard deviations. The first row uses the primary reference in the text. Blue indicates that the 90\% interval lies within the bounds, and red indicates that it extends beyond them (Table~\ref{tab:tost}).}
	\label{fig:framing}
\end{figure}

The Human-label payoff advantage was \pounds0.160 per round under the AI Agent framing and \pounds0.148 under AI Delegation (Fig.~\ref{fig:framing}A and Table~\ref{tab:h2}). The estimated change in the gap under Delegation was small ($b = -0.007$, 95\% CI [$-0.084$, 0.069], $p = 0.848$).

The primary equivalence comparison used the participant-level standard deviation of mean payoff (0.357) to express the preregistered smallest effect size of interest of 0.20 standard deviations in payoff units. The resulting bounds were $\pm 0.071$. The 90\% confidence interval for the interaction [$-0.072$, 0.057] extended just beyond the lower bound. Two one-sided tests (TOST) gave $p = 0.051$. The interval therefore leaves room for a framing effect of the prespecified size, and the equivalence conclusion depends on the choice of reference standard deviation (Fig.~\ref{fig:framing}B and Table~\ref{tab:tost}). The results provide little evidence of the smaller penalty under Delegation predicted in H5. Descriptive first-offer comparisons also showed lower offers to AI opponents under both framings (Supplementary Materials, \textit{Hypothesis 5: AI Agent versus AI Delegation framing}). The gap was present under each description of AI, while uncertainty remains about how much framing changes its size.

\FloatBarrier
\section*{Discussion}

Players presented as AI received smaller payoffs even when all three players agreed to the same deal. This comparison concerns three-player agreements that formed in mixed groups. Across all rounds in mixed groups, AI players earned about 18\% less than Human players. They joined final agreements less often and received smaller payoffs when included. A human played every position under the same rules, so differences in how players were presented could matter even when the underlying capabilities and incentives were the same.

A third player makes it possible to distinguish who reaches an agreement from what each party receives \cite{baron1989,frechette2003,kagel2010}. In earlier experiments using the same bargaining structure, exclusion from coalitions accounted for most of the gender gap in earnings \cite{baranski2024}. Research on partner choice also gives reason to expect counterparts presented as AI to be included less often \cite{jiang2025}. In our experiment, the descriptive decomposition associated most of the payoff gap with the terms received by players who joined an agreement. This distinction matters for AI-mediated representation. A safeguard that tracks who participates in decisions can miss disparities in how gains are distributed.

Group composition also mattered. The payoff gap was larger when Human players were in the majority. In comparisons holding both opponents' labels fixed, the estimated Human-label advantage was positive in all three opponent configurations. This extends work in settings where people face a single counterpart presented as AI \cite{zhou2026,ishowo2019,dvorak2025}. The composition of two-player agreements also varied with the focal player's label. Players saw their opponents' labels while remaining unaware of their own. Outcomes of two-player agreements reflect joint decisions, and the comparisons between focal-player labels also vary in group composition because they are based on players who faced one opponent of each label.

Players offered less to AI opponents in their own first proposal of a round. The points they retained in two-player agreements were slightly higher with AI partners, but this difference was not statistically reliable. For the lower offers, earlier work on self-interest, fairness and mental-state attribution suggests several possible explanations \cite{karpus2021,demelo2014,march2021,niszczota2026,lee2021,swiderska2019,sanfey2003,bonnefon2024arp}. Participants may have anticipated different bargaining behaviour from AI opponents or held them to different standards of fairness. Other social cues attached to AI, including an assigned gender, can also shape responses to it as a cooperation partner or manager \cite{spadaro2023,cui2026managers}. The design isolates responses to being presented as AI while leaving open how much they reflect strategic expectations, perceived moral standing or both. With the same partners across ten rounds, there was little evidence of systematic change in the overall gap. Other studies have found smaller or undetectable label differences when agents act fairly, cooperative norms are shared, or feedback makes individual contributions difficult to attribute \cite{wang2026,mutzner2026,makovi2023}.

The payoff gap appeared when AI players were presented as autonomous agents and when they were presented as delegates for humans. The estimated difference between the framings was close to zero, and the primary equivalence comparison left room for an effect of the prespecified size. In settings where people actually delegate to AI, delegation can also change the decisions themselves \cite{kobie2025nature}. Our experiment isolates the response to how a counterpart was described. That distinction matters as AI agents take on roles in collective decisions \cite{rahwan2019,zhu2026,fernandez2022}, where the identity of a partner and the terms of an agreement can both shape the distribution of gains \cite{jackson2002}.

These findings raise one tension and one open question for human--AI collaboration. Research on cooperation and collective intelligence identifies circumstances in which artificial agents can contribute effectively \cite{crandall2018,cui2024collective}, although performance gains depend on how humans and AI are combined \cite{vaccaro2024}. Such gains may be harder to realise if other people respond unfavourably to AI involvement \cite{ishowo2019,dvorak2025}. A person could therefore face a trade-off between the task benefits of delegation and the social cost of being represented by AI. Our experiment did not compare AI with human performance or observe actual delegation choices, but the payoff gap under the Delegation framing shows why this possibility deserves attention.

The open question concerns the relationship between partner selection and the terms of agreement. Human players more often ended in two-player agreements with Human opponents. We found no clear evidence that the deals they reached with AI partners were worse for them: the points retained were slightly higher with AI partners, but the difference was not statistically reliable. Agreements were reached jointly, and we observe terms only for deals that formed. If such a difference exists, the social appeal of a partner and the bargaining returns available from that partner may diverge. The present data cannot establish this, and it remains a question for designs that manipulate partner choice directly.

The study used a stylised game, ten rounds and a single online sample in the United States. Real AI agents may bargain differently from human participants, and responses to artificial counterparts vary across countries \cite{karpus2025}. The estimates are conditional on retention in the sample. Table~\ref{tab:attrition} reports pooled attrition counts, leaving open whether dropout varied by assigned label and whether that affected the estimated gap. The comparisons within final agreements are conditional on agreements that formed. The decomposition therefore describes how the overall payoff gap is distributed across inclusion and terms in the observed data.


\section*{Materials and Methods}

\subsection*{Participants and recruitment}

Participants were recruited through Prolific \cite{palan2018,peer2017} from a United States-based panel and completed the study online in English. The experiment itself was programmed in oTree \cite{chen2016}. Data were collected over ten sessions. After exclusions, the analytical sample comprised 2,234 participants from 776 groups, originally formed as triads. Demographically, 48.6\% of participants were female and 50.0\% male, with a mean age of 38.3 years (SD = 12.5, range 18 to 88; Table~\ref{tab:demographics}). A valid completion required passing the comprehension quiz, finishing all ten rounds, and passing the manipulation check. The recruitment target and power analysis are reported in the Supplementary Materials.

\subsection*{Design and assignment}

The design crossed four group compositions with two framings of the AI label. Within each group, participants were randomly assigned to one of three positions. Eight position-specific Human/AI label patterns yielded four group compositions (HHH, AHH, AAH and AAA). The eight patterns cycled in a fixed order as groups formed, separately within each framing condition. This produced group compositions in a ratio of approximately 1:3:3:1. The framing condition alternated between successive blocks of three participants in order of arrival. Positions, labels, group membership and framing remained fixed over ten rounds. Fig.~\ref{fig:design}B and Table~\ref{tab:sample} show the number of retained participants and groups in each condition.

\subsection*{The bargaining game}

The game adapts the majoritarian bargaining design of Baranski and colleagues \cite{baranski2024}, an experimental divide-the-dollar game in the tradition of legislative bargaining \cite{baron1989,frechette2003}, by manipulating the composition of bargaining triads. In our case, we replace the gender composition used by Baranski and colleagues \cite{baranski2024} with Human and AI labels. Groups of three bargained over the division of a common surplus for ten consecutive rounds. Within a round, players moved continuously in time on a triangular allocation grid, each cell representing one division of 12 points among the three positions (Fig.~\ref{fig:design}A). Selecting a cell constituted a proposal, and players could revise, counter or withdraw proposals during the round. Players could communicate only through the grid, by making proposals and indicating agreement on screen. Agreement required a majority: any two of the three players on the same cell formed a coalition. Importantly, all three could also agree on a cell, but the coalition only finalised if at least two players held it. A coalition had to be held for ten consecutive seconds to become final, at which point the round ended. A player could break a forming coalition by moving away before that interval elapsed. Each round lasted at most 120 seconds, and a round in which no coalition was finalised ended without payment.

The surplus shrank continuously during each round. Each point carried a value that was highest at the start of the round and decayed linearly to zero over 120 seconds, from \pounds0.255 to nothing, so agreeing sooner was worth more to every member of the eventual coalition than agreeing later. Delay was therefore costly to all parties simultaneously and did not favour any position.
Still, it created tension between holding out for a larger share of the points and preserving the value of the eventual payoff.

\subsection*{AI-label manipulation and framing conditions}

The interface displayed the Human or AI label of each opponent and marked the player's own position as ``You''. Participants were unaware of the labels assigned to their own positions. Under the AI Agent framing, participants were told that their group might include AI agents trained on human gameplay. Under AI Delegation, they were told that some other players might have delegated their decisions to an AI agent trained on human gameplay. The experiment assigned the labels. Participants made every move themselves and had no option to delegate. Opponents' labels remained visible throughout play.

\subsection*{Procedure}

After giving informed consent, participants read instructions covering the allocation grid, the majority rule, the ten-second confirmation interval, the shrinking value of points, and the payment rule, and then completed a comprehension quiz. Only those who passed this comprehension quiz were allowed to continue. They next practised in an unincentivised single-player sandbox, in which the other two positions were simulated, to familiarise themselves with the interface. They then saw the labels of their two opponents and the framing description for their condition, and answered a short question identifying the opponents' labels. Participants then played the ten incentivised rounds with the same partners and the same labels throughout. After the final round, they completed a manipulation check that asked them to identify the labels of their two opponents; participants who failed it were excluded from analysis, and the remainder completed a post-game questionnaire and a final question about whether they had doubted that the opponents shown as AI were AI systems. Of the 2,234 retained participants, 2,227 answered this question, and 204 (9.2\%) said they had. All participants were then debriefed, including an explicit statement that every player was human and that no artificial agent made any decisions.

\subsection*{Payment}

Every participant received a fixed base payment of around \pounds6 per hour for their time, paid through the recruitment platform, regardless of how they bargained. In addition, they could earn a performance-dependent bonus. The bonus was based on one round rather than on cumulative earnings. At the end of the session, one of the ten rounds was randomly selected for each participant. The bonus was the number of points the participant held in the final division of that round, multiplied by the per-point value in pounds at the moment the coalition was finalised, so the bonus could reach at most \pounds3.06 (12 points at \pounds0.255). The current value of a point was shown on screen throughout each round, so participants could see what an agreement was worth while bargaining. Payoffs reported in the Results are in the same units, pounds per round. A participant who was excluded from the coalition in the selected round, or whose group reached no agreement in it, received no bonus for that round. Because the paid round was unknown during bargaining, every round carried the same expected consequence, and neither exclusion from a coalition nor the terms of a division was costless.

\subsection*{Measures}

We recorded moves with timestamps, allowing us to reconstruct bargaining sequences.
The primary outcome was the participant's final payoff in a round, defined as the points held in the finalised division multiplied by the per-point value at the moment of agreement, and as zero for a participant who was not part of the finalised coalition or whose group reached no agreement, as under the payment rule. We also analysed whether the participant was part of the finalised agreement, and restricted payoff to rounds in which they were. From the move sequence, we also derived the allocation offered to each opponent in a participant's first proposal of a round, the identity of the partner in each finalised two-player coalition, the points a participant retained in finalised coalitions, whether a forming coalition broke before confirmation, and the order in which players arrived at a division that all three accepted. The post-game questionnaire recorded self-reported bargaining behaviour, perceptions of the interaction, and, for participants in mixed or all-AI groups, attitudes toward counterparts presented as AI. Questionnaire outcomes are reported in the Supplementary Materials.

\subsection*{Exclusions and analytical sample}

Participants were excluded if they failed the instruction quiz, did not complete all ten rounds, or failed the manipulation check. Of the 3,249 participants who joined a bargaining group, 678 (20.9\%) dropped out via a timeout, the comprehension quiz, or self-exit; 84 (2.6\%) failed the manipulation quiz; and 233 left before completing all ten rounds, 149 because another member of their group dropped out. As an exception, 13 participants who completed the game and the post-game survey but timed out on the final suspicion question were retained, because their round-level data remain valid. Six selected an answer to the suspicion question, which was recorded when time ran out although they did not submit the page, and seven did not answer. A further, non-preregistered data-quality exclusion removed group-rounds whose recorded per-point value exceeded the design maximum of \pounds0.255 because a participant's device clock was out of step with the server (133 group-rounds in 34 groups). Because exclusions were applied to participants and to group-rounds, some retained groups contained fewer than three analysable participants or fewer than ten rounds. Table~\ref{tab:attrition} gives the pooled attrition and exclusion counts. Fig.~\ref{fig:design}B and Table~\ref{tab:sample} show the retained sample by group composition and framing. The Supplementary Materials describe the exclusions and data reconstruction.

\subsection*{Statistical analysis}
We restricted the primary payoff comparison between Human and AI players to mixed-label groups (AHH and AAH).
We tested the hypotheses using linear mixed-effects models, linear regression or generalised estimating equations, depending on the outcome. Models included round and the relevant label and composition terms. Players in a group divide one pot and stay together for ten rounds, so observations within a bargaining group are not independent. Standard errors and confidence intervals therefore treat the bargaining group as the unit of independence throughout. The Supplementary Materials give the model and clustering specification for each outcome. We centred the round at the midpoint so that intercepts describe the middle of the session. For the framing comparison, we also used two one-sided tests against a smallest effect size of interest of $|d| = 0.20$. Apart from these equivalence tests, all tests are two-sided, and confidence intervals are two-sided 95\% intervals unless otherwise specified. For descriptive means and percentages in the main-text figures, we used a cluster bootstrap with 2,000 resamples of bargaining groups. Intervals around model estimates (Figs.~\ref{fig:penalty}C and \ref{fig:framing}B) come from the corresponding models and use the group-clustered standard errors. The Supplementary Materials also detail the equivalence tests, robustness analyses and departures from the preregistered analysis plan. Model coefficients and confidence bounds are generally shown to three decimal places; a displayed bound of 0.000 need not be exactly zero.

\subsection*{Ethics}

The University College Dublin Ethics Board approved the study protocol with reference number 275-LS-C-25-Yasseri. All participants gave informed consent before taking part. The design involved deception, in that participants were told that some positions were, or were controlled by, artificial agents when, in fact, every player was human. No participant was deceived about the payment rule or about the consequences of their own choices, and all participants were fully debriefed at the end of the session about the use of AI labels and the absence of any artificial agent.


\bibliography{references}
\bibliographystyle{sciencemag}


\section*{Acknowledgments}
We thank Nima Hamedani Radja, Siobh\'an Grayson, and Diogo Geraldes for their valuable assistance and feedback.
\paragraph*{Funding:}
This publication has emanated from research conducted with the financial support of
Research Ireland under Grant number IRCLA/2022/3217. TY acknowledges support from Workday Inc.

\paragraph*{Author contributions:}
TG: Conceptualisation, Methodology, Formal analysis, Writing - Original Draft, Writing - Review \& Editing.\newline 
NM: Methodology, Formal analysis, Writing - Original Draft, Writing - Review \& Editing.\newline 
TY: Conceptualisation, Methodology, Resources, Writing - Review \& Editing, Supervision.
\paragraph*{Use of AI-assisted technologies:}
ChatGPT (OpenAI) was used to assist with manuscript revision and editorial review. Claude (Anthropic) was used to assist with reviewing the statistical analysis. The authors take responsibility for the content of the manuscript and the reported analyses.
\paragraph*{Competing interests:}
The authors declare that they have no known competing financial interests or personal relationships that could have appeared to influence the work reported in this paper.
\paragraph*{Data, code and materials availability:}
The accompanying reproducibility package contains de-identified versions of the 10 raw oTree exports, the scripts for all confirmatory and exploratory analyses reported in the manuscript and Supplementary Materials, model outputs, figure-source data, and the scripts used to generate the publication figures. Before deposit, we removed Prolific identifiers, participant and session labels, participant start times, and free-text responses from the exports, and replaced participant and session codes with pseudonyms; analyses run on the de-identified exports reproduce the reported results. We do not include participant demographics from Prolific in the package to protect participants' confidentiality. We computed Tables~\ref{tab:demographics} and \ref{tab:demorobust} from them, and we include the scripts and outputs.
No physical materials were generated in this work.


\newpage


\subsection*{Supplementary Materials}
Supplementary Text\\
Figs. S1 to S5\\
Tables S1 to S16

\newpage


\renewcommand{\thefigure}{S\arabic{figure}}
\renewcommand{\thetable}{S\arabic{table}}
\renewcommand{\theequation}{S\arabic{equation}}
\renewcommand{\thepage}{S\arabic{page}}
\setcounter{figure}{0}
\setcounter{table}{0}
\setcounter{equation}{0}
\setcounter{page}{1}
\linespread{1}\selectfont

\begin{center}
\section*{Supplementary Materials for\\ \scititle}

Tahmineh~Godazgar,
Nico~Mutzner,
Taha~Yasseri\\
\end{center}

\subsubsection*{This PDF file includes:}
Supplementary Text\\
Figures S1 to S5\\
Tables S1 to S16

\newpage

This supplement reports sample construction, model specifications, expanded descriptive results, robustness analyses, three-player-agreement diagnostics, proposal success, coalition instability, timing, and survey outcomes. Human and AI refer to the roles shown to opponents. Confirmatory tests are separated from exploratory and descriptive analyses throughout.

\subsection*{Supplementary Methods}

\subsubsection*{Sample construction and exclusions}

Recruitment continued until we reached the preregistered target of 2,000 valid completions. We set the target based on a power analysis showing that 2,000 participants yield approximately 666 independent bargaining triads, providing 90\% power at $\alpha = 0.05$ to detect the smallest effect size of interest, $d = 0.20$. This target also supports equivalence testing against the same bound, used for the framing comparison in Hypothesis 5.

The per-point exclusion described in the main text arises as follows. The per-point value was not computed on the server: each participant's browser counted it down from \pounds0.255 using its own clock and the round-start time set by the server, and when the round ended, the server stored the last value it received for the whole group and used it for payment. A device clock running behind the server therefore produced stored values above \pounds0.255. All such values lie exactly on the grid the countdown can produce, with a constant implied clock lag within a group, identifying them as clock offsets rather than data corruption. Because the recorded value, and hence the payoff, is wrong in such a round, the whole group-round was excluded for all three members. Smaller offsets that keep the value below the maximum cannot be detected this way; judged against the server's round timestamps, they affect about 4\% of retained group-rounds.

Table~\ref{tab:attrition} gives the pooled attrition and exclusion counts. Of 5,475 participants who entered the study, 2,226 never joined a group in the synchronous three-player matching procedure. The listed dropouts and manipulation-check exclusions reduce participants from 3,249 group entrants to 2,254. Excluding 26 participants with missing post-game survey data and six whose recorded per-point value exceeded the maximum in all rounds leaves 2,222. The final sample adds 13 participants who completed the game and survey but timed out on the suspicion question and were recorded as dropouts. One completer who never made a move contributed no observations. These adjustments yielded 2,234 participants. Table~\ref{tab:sample} shows the retained participants and groups by composition and framing.

Table~\ref{tab:demographics} describes the analytical sample. Demographic information comes from the participants' Prolific profiles and was matched through the Prolific ID entered at the start of the study. Sex was recorded as female or male for 2,204 of the 2,234 participants and age for 2,223. Values are missing where Prolific no longer returned them or, for three participants, where the ID entered could not be matched to a Prolific record. Student and employment status were unavailable for about 14\% of participants. The composition was similar among participants presented as Human and as AI.

\begin{table}
	\centering
	\caption{\textbf{Participant attrition and exclusions.} Counts pooled across the ten data-collection sessions and assigned labels. Indented rows show the exclusions and adjustments applied after each subtotal.}
	\label{tab:attrition}
	\small
	\begin{tabular}{lr}
		\\
		\hline
		\textbf{Stage} & \textbf{Participants}\\
		\hline
		Entered the study & 5,475\\
		\quad Never joined a bargaining group & 2,226\\
		\hline
		Joined a bargaining group & 3,249\\
		\quad Dropped out: timeout, comprehension quiz, or self-exit & 678\\
		\quad Excluded: failed the manipulation check & 84\\
		\quad Dropped out: another group member left & 149\\
		\quad Left the game or disconnected & 66\\
		\quad Left the game or disconnected (page not recorded) & 18\\
		\hline
		Remaining after the preceding exclusions & 2,254\\
		\quad Excluded: post-game survey data absent & 26\\
		\quad Excluded: every round's per-point value above the maximum & 6\\
		\hline
		Remaining after survey and per-point exclusions & 2,222\\
		\quad Added: completed game and survey, timed out on the final question (counted above) & 13\\
		\quad Omitted: completer who never made a move & 1\\
		\hline
		Analytical sample & 2,234\\
		\hline
	\end{tabular}
\end{table}

Ten oTree exports were processed into one row per participant move. Group-rounds above the per-point maximum (see above) and malformed cell-history spillover rows were excluded during reconstruction. Groups were identified by session and group number, because files holding several sessions reuse group numbers. The clean move-level dataset contained 70,955 rows, of which 66,017 passed the manipulation-check filter. Because exclusions were applied to participants and to group-rounds, some retained groups contained fewer than three analysable participants (89 groups) or fewer than ten rounds (35 groups); group counts therefore do not equal the participant count divided by three.

\begin{table}
	\centering
	\caption{\textbf{Analytical sample by group composition and AI framing.} Because exclusions were applied to participants and to group-rounds, retained participant counts do not equal three times the number of groups.}
	\label{tab:sample}
	\small
	\begin{tabular}{lrrrrrr}
		\\
		\hline
		& \multicolumn{2}{c}{\textbf{AI Agent}} & \multicolumn{2}{c}{\textbf{AI Delegation}} & \multicolumn{2}{c}{\textbf{Total}}\\
		\textbf{Group composition} & Participants & Groups & Participants & Groups & Participants & Groups\\
		\hline
		HHH (all Human) & 132 & 46 & 135 & 47 & 267 & 93\\
		AHH (one AI, two Human) & 410 & 142 & 422 & 147 & 832 & 289\\
		AAH (two AI, one Human) & 442 & 153 & 438 & 153 & 880 & 306\\
		AAA (all AI) & 123 & 43 & 132 & 45 & 255 & 88\\
		\hline
		Total & 1,107 & 384 & 1,127 & 392 & 2,234 & 776\\
		\hline
	\end{tabular}
\end{table}

\subsubsection*{Outcome construction and inferential models}

Final payoff equalled raw final points multiplied by the remaining-money factor. It was set to zero in participant-rounds in which the participant was not part of the finalised coalition (2,969 participant-rounds) or no agreement was reached (60). Hypothesis 1 models used participant-round observations and a group random intercept, with standard errors clustered by bargaining group. The variance of the group random intercept was estimated at zero in the models that count rounds without inclusion as zero. Hypothesis 2 used logistic models fitted with generalised estimating equations (GEE) on one finalised two-player agreement per participant-round, with an independence working correlation and standard errors clustered by bargaining group. Hypothesis 3 expanded each first proposal into two opponent-allocation observations and used participant random intercepts, with standard errors clustered by bargaining group. The Hypothesis 4 analysis retained one finalised two-player agreement per participant-round and used a participant random intercept, with standard errors clustered by bargaining group. Hypothesis 5 used round-level observations with group-clustered standard errors. Arrival-position and payoff models for three-player agreements used group-clustered standard errors, and binary outcomes for three-player agreements used GEE clustered by bargaining group. For the mixed-effects models of Hypotheses 1, 3 and 4, confidence intervals and $p$-values use a $t$ distribution with the number of bargaining groups minus one as degrees of freedom. The GEE models, the Hypothesis 5 model, and the models for three-player agreements use the normal distribution. The means in Tables~\ref{tab:h1means}, \ref{tab:h2}, \ref{tab:grandpay}, and \ref{tab:timing} and the error bars in the supplementary figures use group-clustered standard errors and the same $t$ distribution. Round was centred at 5.5 in interaction models.

In the main-text figures, error bars and shaded bands around means and percentages are 95\% confidence intervals (CIs) from a cluster bootstrap that resamples bargaining groups with replacement (2,000 resamples), so that every descriptive interval treats the group as the unit of independence. Intervals around model estimates are taken from the corresponding models.

\subsubsection*{Preregistration and reporting decisions}

Hypotheses 1 to 5 correspond to the preregistered hypotheses H1, SH1, SH2, SH3, and H2, respectively (\url{https://osf.io/w6xfd}); the model identifiers in the tables below retain the preregistered labels. The preregistration specified directional, one-sided tests at $\alpha = 0.05$. We report two-sided tests and two-sided 95\% confidence intervals throughout, which is the stricter criterion, and use 90\% intervals for the equivalence tests. Under the preregistered one-sided criterion, the Hypothesis 4 estimate for retained points would be significant (one-sided $p = 0.044$) and the estimate for discounted payoff would not (one-sided $p = 0.076$). The five hypotheses were tested without a multiple-comparison correction.

Standard errors clustered by bargaining group were not specified in the preregistration. For Hypothesis 2, the analysis used group-clustered GEE logistic models on one observation per participant-round rather than the mixed-effects logistic model specified in the preregistration. The preregistration described the Hypothesis 1 test as a between-participants comparison of one total payoff per participant, whereas the reported model uses participant-round observations. We also report the participant-level comparison (Table~\ref{tab:h1means}). It leads to the same conclusion: the mean difference was \pounds0.154 per round ($t(594) = 7.79$, $p < 0.001$, with standard errors clustered by bargaining group; Cohen's $d = 0.44$). The preregistration listed payoff, coalition choice, and scores as outcomes for the framing hypothesis.
Framing-specific first-offer comparisons are reported descriptively, without a separate moderation test. We do not report framing-moderation tests for partner choice or points retained.

The null result for the framing interaction (Hypothesis 5) was additionally assessed using two one-sided tests (TOST) against the preregistered smallest effect size of interest of $|d| = 0.20$. Because that bound was specified for the primary between-participant comparisons rather than for an interaction, it was scaled to the participant-level standard deviation of mean payoff in mixed groups ($SD = 0.357$, giving bounds of $\pm 0.071$ payoff units).

Coalition instability was preregistered as an exploratory analysis. Proposal success, the formation of three-player agreements, arrival order, and allocation within three-player agreements were not preregistered. Three-player-agreement mechanisms, proposal success, instability, arrival sequences, and survey analyses are treated as exploratory unless they directly test a stated hypothesis.

\subsection*{Supplementary Results}

\subsubsection*{Analytical and reporting details}

The unit of observation differed across analyses: participant-level means were used for the descriptive overall payoff comparison, participant-round observations for outcomes measured within a round, focal-player observations in finalised coalitions for partner choice, and two opponent-allocation observations per first proposal. The mixed-effects payoff models used participant-round observations. All model intervals use standard errors clustered by bargaining group. Preregistration status, test direction, reporting conventions and model specifications are described in Supplementary Methods.

\subsubsection*{Hypothesis 1: payoff disadvantage and robustness}

\begin{figure}
	\centering
	\includegraphics[width=1.0\textwidth]{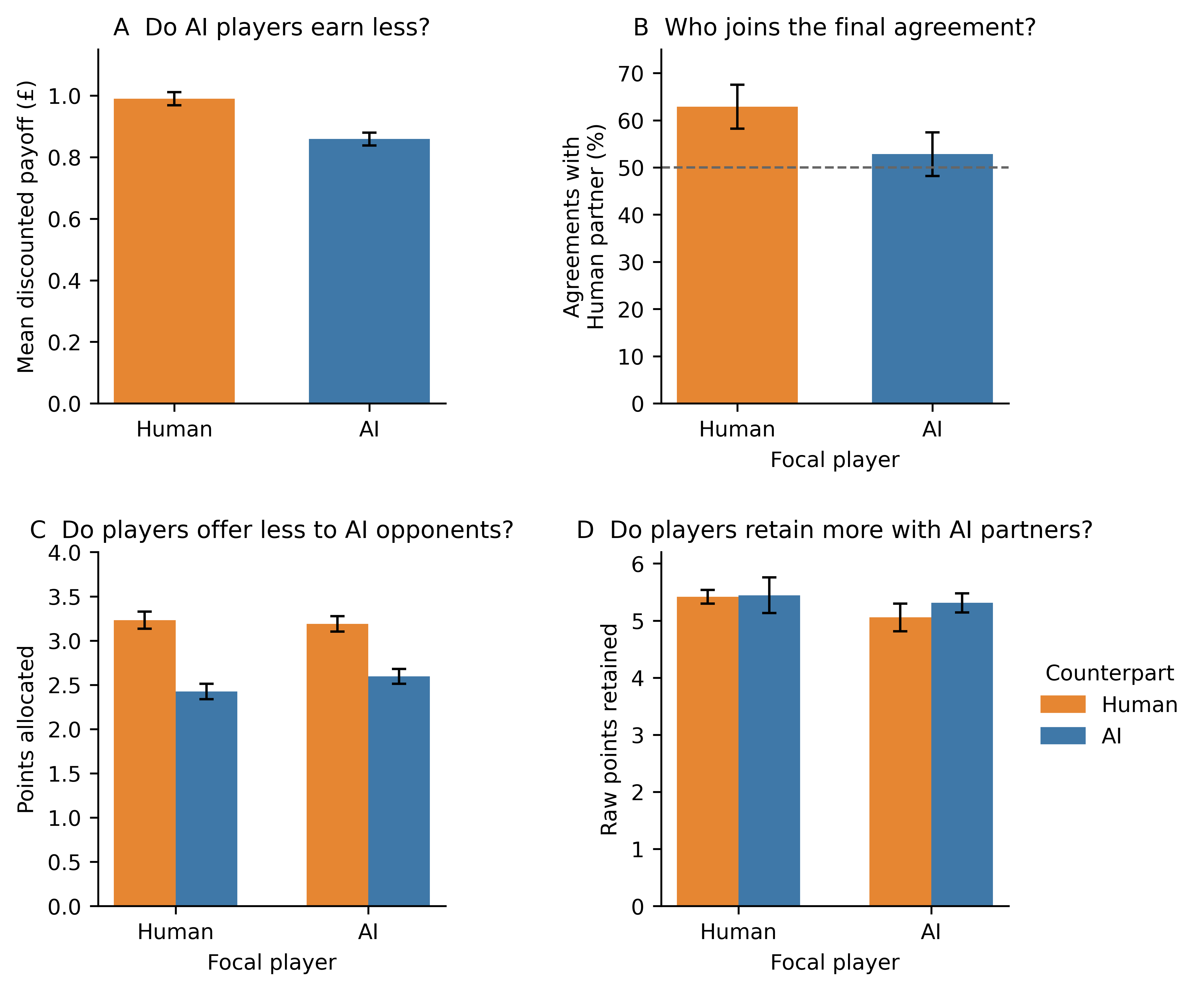}
	\caption{\textbf{Descriptive results by focal player label.} These comparisons complement Figs.~\ref{fig:penalty} and \ref{fig:routes}. A focal player faces one opponent of each label only in AHH groups (Human focal player) or AAH groups (AI focal player). The comparisons in (B) and (D) therefore also vary in group composition. (\textbf{A}) Participant-level mean discounted final payoff (\pounds) by player label in mixed groups, in rounds in which the player was part of the agreement. (\textbf{B}) Percentage of finalised two-player agreements containing a Human partner; the dashed line marks 50\%. (\textbf{C}) Mean first-proposal allocation by proposer and opponent label across all group compositions. (\textbf{D}) Mean raw points retained in finalised two-player agreements for focal players who faced one opponent of each label. Error bars show 95\% confidence intervals from standard errors clustered by bargaining group.}
	\label{fig:slabels}
\end{figure}

Figure~\ref{fig:slabels} summarises descriptive comparisons by focal-player label.

\begin{figure}
	\centering
	\includegraphics[width=1.0\textwidth]{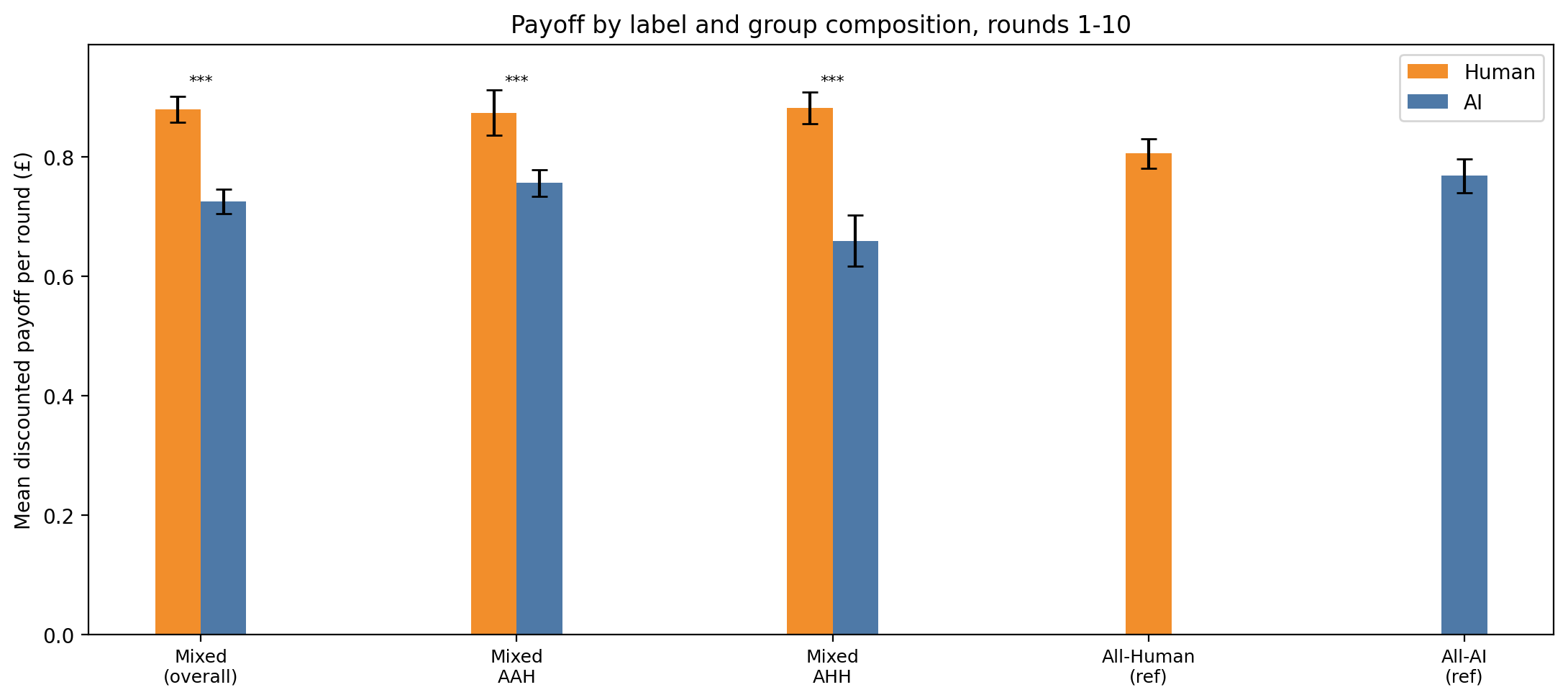}
	\caption{\textbf{Final payoff by label and group composition across rounds 1-10.} Rounds in which the participant was not part of the agreement count as zero. Error bars show 95\% confidence intervals from standard errors clustered by bargaining group; significance marks are pairwise comparisons of Human and AI players with group-clustered standard errors (*** $p < 0.001$).}
	\label{fig:spayoff}
\end{figure}

The descriptive Human-label payoff advantage was \pounds0.117 per round in AAH groups and \pounds0.223 in AHH groups (Table~\ref{tab:h1means}). In AAH groups, the gap was smaller in rounds 6--10 ($d = 0.192$) than across all ten rounds ($d = 0.345$). In AHH groups, the corresponding estimates were $d = 0.625$ and $d = 0.590$. These round-window comparisons are descriptive (Fig.~\ref{fig:penalty}B). Fig.~\ref {fig:spayoff} shows payoff comparisons by label and group composition. Across mixed groups, Human and AI players averaged 4.190 and 3.505 raw final points per round, respectively. In rounds in which the player joined the final agreement, the mean per-point values were \pounds0.209 and \pounds0.207, against a starting value of \pounds0.255.

Table~\ref{tab:context} reports exploratory models of group and opponent composition. In the mixed-group model, AHH membership was associated with a lower payoff for AI players ($b = -0.102$, 95\% CI [$-0.150$, $-0.054$], $p < 0.001$). In the model spanning all four group compositions, the Human-label main effect is the contrast against two AI opponents. Adding the relevant Human-label interaction gives the contrast against one opponent of each label or against two Human opponents. Table~\ref{tab:exclusion} reports the descriptive decomposition of the payoff gap and the estimates for all rounds and for rounds with inclusion.

\begin{table}
	\centering
	\caption{\textbf{Participant-level mean discounted payoff across rounds 1-10.} Rounds in which the participant was not part of the finalised coalition count as zero. Scope refers to the group compositions included in each comparison. Confidence intervals use standard errors clustered by bargaining group. Payoffs and payoff differences are in pounds per round.}
	\label{tab:h1means}
	\small
	\begin{tabular}{llrrl}
		\\
		\hline
		\textbf{Scope} & \textbf{Label} & \textbf{N} & \textbf{Mean payoff} & \textbf{95\% CI}\\
		\hline
		mixed\_overall & Human & 849 & 0.879 & [0.858, 0.901]\\
		mixed\_overall & AI & 863 & 0.725 & [0.705, 0.746]\\
		mixed\_AAH & Human & 293 & 0.873 & [0.835, 0.911]\\
		mixed\_AAH & AI & 587 & 0.756 & [0.734, 0.778]\\
		mixed\_AHH & Human & 556 & 0.882 & [0.856, 0.909]\\
		mixed\_AHH & AI & 276 & 0.659 & [0.616, 0.702]\\
		all\_human & Human & 267 & 0.805 & [0.780, 0.830]\\
		all\_ai & AI & 255 & 0.768 & [0.740, 0.797]\\
		\hline
	\end{tabular}
\end{table}
\begin{table}
	\centering
	\caption{\textbf{Exploratory models of group and opponent composition.} Neither model
		was preregistered. Both are linear mixed-effects models of discounted payoff (zero in rounds
		without inclusion in the finalised coalition) with a group random intercept and round centred
		at its midpoint. Standard errors are clustered by bargaining group. In the first model, the coefficient for AHH group is conditional on the AI label,
		so it gives the effect of AHH membership for AI players. In the second, the interaction
		terms test whether the opponent-composition effect differs by the focal player's own label.
		The bottom block gives the Human-AI difference for each opponent combination, the Human-label
		main effect plus the relevant interaction, with confidence intervals from the model's
		group-clustered covariance matrix.}
	\label{tab:context}
	\small
	\begin{tabular}{lrll}
		\\
		\hline
		\textbf{Term} & \textbf{b} & \textbf{95\% CI} & \textbf{p}\\
		\hline
		\multicolumn{4}{l}{\textit{Group composition (mixed groups only; reference AAH), n = 16,734}}\\
		Intercept & 0.762 & [0.740, 0.783] & \ensuremath{<}0.001\\
		Human label & 0.110 & [0.057, 0.164] & \ensuremath{<}0.001\\
		AHH group & $-$0.102 & [$-$0.150, $-$0.054] & \ensuremath{<}0.001\\
		Human label $\times$ AHH group & 0.113 & [0.029, 0.197] & 0.009\\
		Round (centred) & 0.013 & [0.011, 0.014] & \ensuremath{<}0.001\\
		\hline
		\multicolumn{4}{l}{\textit{Opponent composition (all groups; reference two AI opponents), n = 21,854}}\\
		Intercept & 0.772 & [0.749, 0.796] & \ensuremath{<}0.001\\
		Human label & 0.100 & [0.056, 0.144] & \ensuremath{<}0.001\\
		Opponents: one of each label & $-$0.011 & [$-$0.042, 0.021] & 0.501\\
		Human label $\times$ one of each & 0.021 & [$-$0.024, 0.067] & 0.353\\
		Opponents: both Human & $-$0.113 & [$-$0.161, $-$0.064] & \ensuremath{<}0.001\\
		Human label $\times$ both Human & 0.047 & [$-$0.019, 0.113] & 0.163\\
		Round (centred) & 0.012 & [0.011, 0.014] & \ensuremath{<}0.001\\
		\hline
		\multicolumn{4}{l}{\textit{Human advantage by opponent combination (from the second model)}}\\
		Two AI opponents & 0.100 & [0.056, 0.144] & \ensuremath{<}0.001\\
		One opponent of each label & 0.121 & [0.087, 0.155] & \ensuremath{<}0.001\\
		Two Human opponents & 0.147 & [0.097, 0.196] & \ensuremath{<}0.001\\
		\hline
	\end{tabular}
\end{table}
\begin{table}
	\centering
	\caption{\textbf{Exclusion and terms of inclusion in the Hypothesis 1 payoff gap.} Top (descriptive): share of participant-rounds in which the participant was part of the finalised agreement, mean payoff in those rounds, and a symmetric decomposition of the per-round gap in expected payoff, $P(\text{included}) \times E[\text{payoff} \mid \text{included}]$, into a part due to inclusion (exclusion) and a part due to payoff when included (terms). All top-panel quantities weight each participant-round equally, so that the decomposition is exact. The gaps therefore differ slightly from the participant-level mean differences in Table~\ref{tab:h1means} (0.154 in mixed and 0.117 in AAH groups). Inclusion by label in mixed groups: GEE logit b = 0.404, 95\% CI [0.244, 0.563], p \ensuremath{<} 0.001. Bottom (model estimates): the Hypothesis 1 mixed-effects estimates (95\% CI, with standard errors clustered by bargaining group) with rounds without inclusion counted as zero, as in the main text, and restricted to rounds in which the participant was included. The last row gives participant-level means. Payoffs and payoff differences are in pounds per round.}
	\label{tab:exclusion}
	\small
	\begin{tabular}{llllll}
		\\
		\hline
		\multicolumn{6}{l}{\textit{Descriptive, per participant-round}}\\
		\textbf{Groups} & \shortstack[l]{\textbf{Included}\\\textbf{(H / AI)}} & \shortstack[l]{\textbf{Payoff if}\\\textbf{included (H / AI)}} & \textbf{Gap} & \textbf{Exclusion} & \textbf{Terms}\\
		Mixed (AHH and AAH) & 88.2\% / 83.5\% & 0.996 / 0.873 & 0.150 & 0.044 (29\%) & 0.106 (71\%)\\
		AHH & 89.0\% / 81.0\% & 0.991 / 0.814 & 0.223 & 0.072 (32\%) & 0.151 (68\%)\\
		AAH & 86.7\% / 84.7\% & 1.005 / 0.899 & 0.110 & 0.019 (18\%) & 0.091 (82\%)\\
		\hline
		\multicolumn{6}{l}{\textit{Hypothesis 1 models}}\\
		\textbf{Term} & \multicolumn{2}{l}{\textbf{Exclusion as zero}} & \multicolumn{3}{l}{\textbf{Included rounds only}}\\
		Human label (H1a) & \multicolumn{2}{l}{0.150 [0.112, 0.188]} & \multicolumn{3}{l}{0.126 [0.089, 0.163]}\\
		Human label $\times$ round (H1b) & \multicolumn{2}{l}{$-$0.004 [$-$0.012, 0.004]} & \multicolumn{3}{l}{$-$0.002 [$-$0.009, 0.005]}\\
		Human label $\times$ AHH group & \multicolumn{2}{l}{0.113 [0.029, 0.197]} & \multicolumn{3}{l}{0.069 [$-$0.010, 0.149]}\\
		Both opponents Human & \multicolumn{2}{l}{$-$0.113 [$-$0.161, $-$0.064]} & \multicolumn{3}{l}{$-$0.084 [$-$0.133, $-$0.036]}\\
		\shortstack[l]{Mean payoff, H / AI\\(mixed; per participant)} & \multicolumn{2}{l}{0.879 / 0.725} & \multicolumn{3}{l}{0.990 / 0.859}\\
		\hline
	\end{tabular}
\end{table}

\subsubsection*{Hypothesis 2: partner selection}

The confirmatory analysis used finalised two-player agreements to identify which opponent ultimately joined the deal. The models reported in Results account for repeated observations within the same bargaining group.

In the group-clustered GEE, the estimated proportion of final agreements that paired the player with a Human opponent declined with round among AI focal players ($b = -0.048$, 95\% CI [$-0.088$, $-0.008$], $p = 0.020$). Among Human focal players, the estimate showed little evidence of change across rounds ($b = 0.004$, 95\% CI [$-0.040$, 0.048], $p = 0.868$). The focal-label by round interaction was $b = 0.052$ (95\% CI [$-0.008$, 0.111], $p = 0.090$, Table~\ref{tab:sh1}).

\begin{table}
	\centering
	\caption{\textbf{GEE logistic models of whether a finalised two-player agreement included the Human opponent.} Positive coefficients indicate higher odds of an agreement with the Human opponent. The models use one observation per participant-round and an independence working correlation. Standard errors are clustered by bargaining group. Round is centred at 5.5.}
	\label{tab:sh1}
	\small
	\begin{tabular}{llrll}
		\\
		\hline
		\textbf{Model} & \textbf{Term} & \textbf{b} & \textbf{95\% CI} & \textbf{p}\\
		\hline
		SH1a\_main & Intercept & 0.108 & [$-$0.077, 0.293] & 0.254\\
		SH1a\_main & Human\_ego & 0.415 & [0.144, 0.686] & 0.003\\
		SH1a\_main & round\_c & $-$0.023 & [$-$0.053, 0.007] & 0.138\\
		SH1b\_interaction & Intercept & 0.102 & [$-$0.084, 0.288] & 0.282\\
		SH1b\_interaction & Human\_ego & 0.424 & [0.151, 0.697] & 0.002\\
		SH1b\_interaction & round\_c & $-$0.048 & [$-$0.088, $-$0.008] & 0.020\\
		SH1b\_interaction & Human\_ego:round\_c & 0.052 & [$-$0.008, 0.111] & 0.090\\
		SH1c\_human\_ego & Intercept & 0.526 & [0.327, 0.726] & \ensuremath{<}0.001\\
		SH1c\_human\_ego & round\_c & 0.004 & [$-$0.040, 0.048] & 0.868\\
		SH1d\_ai\_ego & Intercept & 0.102 & [$-$0.084, 0.288] & 0.282\\
		SH1d\_ai\_ego & round\_c & $-$0.048 & [$-$0.088, $-$0.008] & 0.020\\
		\hline
	\end{tabular}
\end{table}

\subsubsection*{Hypothesis 3: first-offer allocation}
The pooled premium for Human opponents in each player's first proposal in a round was 0.728 points. When both opponents carried the same label, Human proposers offered 3.018 points to Human opponents and 2.501 points to AI opponents in the respective groups. The corresponding values for AI proposers were 3.092 and 2.628 points (Fig.~\ref{fig:routes}B). In the separate models by proposer label, the premium increased across rounds for Human proposers ($b = 0.054$, 95\% CI [0.018, 0.089], $p = 0.003$). The estimate for AI proposers was $b = -0.026$ (95\% CI [$-0.058$, 0.006], $p = 0.116$, Table~\ref{tab:sh2}).

\begin{table}
	\centering
	\caption{\textbf{Mixed-effects models of points allocated to opponents in first proposals.} Positive coefficients indicate more points allocated to Human opponents. Standard errors are clustered by bargaining group. Round is centred at 5.5.}
	\label{tab:sh2}
	\small
	\begin{tabular}{llrll}
		\\
		\hline
		\textbf{Model} & \textbf{Term} & \textbf{b} & \textbf{95\% CI} & \textbf{p}\\
		\hline
		SH2a\_main & opp\_is\_human & 0.727 & [0.609, 0.844] & \ensuremath{<}0.001\\
		SH2b\_opp\_round & opp\_is\_human & 0.728 & [0.609, 0.847] & \ensuremath{<}0.001\\
		SH2b\_opp\_round & opp\_is\_human:round\_c & 0.011 & [$-$0.013, 0.035] & 0.380\\
		SH2c\_human\_ego & opp\_is\_human & 0.867 & [0.682, 1.051] & \ensuremath{<}0.001\\
		SH2c\_human\_ego & opp\_is\_human:round\_c & 0.054 & [0.018, 0.089] & 0.003\\
		SH2d\_ai\_ego & opp\_is\_human & 0.607 & [0.443, 0.770] & \ensuremath{<}0.001\\
		SH2d\_ai\_ego & opp\_is\_human:round\_c & $-$0.026 & [$-$0.058, 0.006] & 0.116\\
		\hline
	\end{tabular}
\end{table}

\subsubsection*{Hypothesis 4: points and payoffs in two-player agreements}

Because an outcome repeats on each move row, the analysis retains one observation per participant and round. It is restricted to finalised two-player agreements for focal players who faced one opponent of each label. The preregistered specification pools focal players across labels. In a model adjusting for focal-player label, the estimated difference in discounted payoff by partner label was \pounds0.038 lower with a Human partner (95\% CI [$-0.087$, 0.011], $p = 0.129$). The same model estimated a payoff that was \pounds0.067 higher when the focal player carried a Human label (95\% CI [0.020, 0.114], $p = 0.005$). The interaction between partner and focal-player label was $b = 0.020$ (95\% CI [$-0.079$, 0.118], $p = 0.693$). The models split by focal-player label are exploratory departures from the preregistered pooled specification (Table~\ref{tab:sh3}).

\begin{table}
	\centering
	\caption{\textbf{Hypothesis 4 mixed-effects models of points and payoffs retained in finalised agreements.} The first two rows give the preregistered pooled specification. Models split by focal-player label are exploratory. Negative coefficients indicate greater retained value with AI partners. One finalised two-player agreement per participant-round. Standard errors are clustered by bargaining group. Payoffs are in pounds per round; points are undiscounted allocations.}
	\label{tab:sh3}
	\small
	\begin{tabular}{lrrll}
		\\
		\hline
		\textbf{Outcome/model} & \textbf{Observations} & \textbf{b} & \textbf{95\% CI} & \textbf{p}\\
		\hline
		Pooled (preregistered): ego\_payoff\_final & 3495 & $-$0.035 & [$-$0.083, 0.013] & 0.152\\
		Pooled (preregistered): ego\_points\_final & 3495 & $-$0.179 & [$-$0.386, 0.027] & 0.089\\
		Pooled, own label adjusted: ego\_payoff\_final & 3495 & $-$0.038 & [$-$0.087, 0.011] & 0.129\\
		Partner label $\times$ focal label & 3495 & 0.020 & [$-$0.079, 0.118] & 0.693\\
		SH3 Human ego: ego\_payoff\_final & 1774 & $-$0.024 & [$-$0.100, 0.053] & 0.546\\
		SH3 Human ego: ego\_points\_final & 1774 & $-$0.117 & [$-$0.442, 0.208] & 0.479\\
		SH3 AI ego: ego\_payoff\_final & 1721 & $-$0.046 & [$-$0.107, 0.015] & 0.137\\
		SH3 AI ego: ego\_points\_final & 1721 & $-$0.240 & [$-$0.509, 0.029] & 0.080\\
		\hline
	\end{tabular}
\end{table}

\subsubsection*{Hypothesis 5: AI Agent versus AI Delegation framing}

The observed payoff gap appeared under both framings (Table~\ref{tab:h2}). The interaction estimate was small and uncertain ($b = -0.007$, 95\% CI [$-0.084$, 0.069], $p = 0.848$). Table~\ref{tab:tost} shows how the equivalence comparison depends on the reference standard deviation. The preregistration listed payoff, coalition choice and scores as outcomes for the framing hypothesis.
We estimated moderation by framing for payoff. The first-offer comparisons by framing are descriptive. We do not report framing-moderation tests for partner choice or points retained, departing from the preregistered analysis plan. These descriptive first-offer means include all group compositions.

Under the AI Agent framing, Human proposers offered 3.296 and 2.365 points to Human and AI opponents, respectively. AI proposers offered 3.156 and 2.554 points. Under AI Delegation, Human proposers offered 3.168 and 2.486 points, and AI proposers offered 3.221 and 2.639 points.

\begin{table}
	\centering
	\caption{\textbf{Descriptive payoff estimates by AI framing and player label.} Mixed-composition groups only; rounds in which the participant was not part of the finalised coalition count as zero. Confidence intervals use standard errors clustered by bargaining group. Payoffs and payoff differences are in pounds per round.}
	\label{tab:h2}
	\small
	\begin{tabular}{llrrl}
		\\
		\hline
		\textbf{Framing} & \textbf{Label} & \textbf{N} & \textbf{Mean payoff} & \textbf{95\% CI}\\
		\hline
		AI Agent & Human & 418 & 0.881 & [0.851, 0.911]\\
		AI Agent & AI & 434 & 0.721 & [0.691, 0.751]\\
		AI Delegation & Human & 431 & 0.878 & [0.846, 0.909]\\
		AI Delegation & AI & 429 & 0.730 & [0.701, 0.758]\\
		\hline
	\end{tabular}
\end{table}

\begin{table}
    \centering
    \caption{\textbf{Sensitivity of the Hypothesis 5 equivalence test to the standardisation.} The preregistration specified a smallest effect size of interest of $|d| = 0.20$ for the primary between-participant comparisons. Applying it to the framing interaction requires a reference standard deviation; the first row is the one used in the main text. All rows use payoff with rounds without inclusion counted as zero and the same estimated interaction ($b = -0.007$, 90\% CI [$-0.072$, 0.057]).}
    \label{tab:tost}
    \small
    \begin{tabular}{lrrr}
        \hline
        \textbf{Reference standard deviation} & \textbf{SD} & \textbf{Bound} & \textbf{TOST \emph{p}}\\
        \hline
        Participant-level mean payoff, mixed groups (used in the main text) & 0.357 & $\pm$0.071 & 0.051\\
        Round-level payoff, mixed groups (the model's own outcome) & 0.591 & $\pm$0.118 & 0.002\\
        Participant-level mean payoff, Human only & 0.345 & $\pm$0.069 & 0.057\\
        Participant-level mean payoff, AI only & 0.353 & $\pm$0.071 & 0.053\\
        \hline
    \end{tabular}
\end{table}

\subsubsection*{Three-player agreements: formation, allocation, and timing}

Three-player agreements were finalised agreements containing all three players. The analysis used rounds with complete move histories for three players who passed the manipulation check. We report frequencies and payoffs within three-player agreements in Tables~\ref{tab:grandprev} and \ref{tab:grandpay}. We defined the first arriver as the player whose final move reached the eventual agreement cell first. A Human player arrived first in 66.9\% of AHH three-player agreements, and an AI player arrived first in 70.2\% of AAH three-player agreements. Both figures are close to the two-thirds benchmark for the majority label (Fig.~\ref{fig:grand}B). The most common sequences were Human--Human--AI in AHH groups (37.3\%) and AI--AI--Human in AAH groups (36.2\%). In a group-clustered GEE, the estimated association between the AI label and joining last was $b = 0.033$ (95\% CI [$-0.137$, 0.204], $p = 0.702$). Across 3,838 reconstructed arrivals, the third player arrived a mean of 4.781 s after the second (Table~\ref{tab:timing}).

\begin{table}
	\centering
	\caption{\textbf{Prevalence of three-player agreements overall and by group composition.} Restricted to coalition rounds with complete manipulation-passed histories for all three players.}
	\label{tab:grandprev}
	\small
	\begin{tabular}{lrrr}
		\\
		\hline
		\textbf{Composition} & \textbf{Coalition rounds} & \textbf{Three-player agreements} & \textbf{Percentage}\\
		\hline
		OVERALL & 6638 & 3838 & 57.8\%\\
		HHH & 778 & 503 & 64.7\%\\
		AHH & 2449 & 1436 & 58.6\%\\
		AAH & 2648 & 1464 & 55.3\%\\
		AAA & 763 & 435 & 57.0\%\\
		\hline
	\end{tabular}
\end{table}

In the composition GEE, AAH groups had a lower prevalence of three-player agreements than HHH groups (b = $-$0.395, 95\% CI [$-$0.675, $-$0.115], p = 0.006); AHH (b = $-$0.250, p = 0.087) and AAA groups (b = $-$0.319, p = 0.051) did not differ reliably from HHH. The prevalence of three-player agreements increased across rounds (b = 0.061, 95\% CI [0.043, 0.079], p \ensuremath{<} 0.001).

\begin{table}
	\centering
	\caption{\textbf{Discounted payoff within three-player agreements.} Participant-round observations in finalised three-player agreements. Confidence intervals use standard errors clustered by bargaining group. Payoffs and payoff differences are in pounds per round.}
	\label{tab:grandpay}
	\small
	\begin{tabular}{llrrl}
		\\
		\hline
		\textbf{Composition} & \textbf{Label} & \textbf{N} & \textbf{Mean payoff} & \textbf{95\% CI}\\
		\hline
		HHH & Human & 1509 & 0.852 & [0.827, 0.878]\\
		AHH & Human & 2872 & 0.904 & [0.877, 0.931]\\
		AHH & AI & 1436 & 0.755 & [0.707, 0.802]\\
		AAH & Human & 1464 & 0.888 & [0.851, 0.925]\\
		AAH & AI & 2928 & 0.803 & [0.782, 0.825]\\
		AAA & AI & 1305 & 0.822 & [0.803, 0.840]\\
		\hline
	\end{tabular}
\end{table}

Controlling for group composition and round, the estimated Human-label payoff advantage was \pounds0.117 (95\% CI [0.073, 0.161], $p < 0.001$). The $p$-value for the Human-label by AHH interaction was 0.152.

Arrival order was associated with payoffs. Relative to first arrivals, second arrivals received 1.007 fewer raw points (95\% CI [$-1.230$, $-0.784$], $p < 0.001$), and final joiners received 2.591 fewer (95\% CI [$-2.776$, $-2.405$], $p < 0.001$). After accounting for arrival position, AI players received 0.554 fewer points (95\% CI [$-0.742$, $-0.365$], $p < 0.001$). The $ p$-value for the AI label by final-joiner interaction was 0.132.

\begin{table}
    \centering
    \caption{\textbf{Third-arrival timing within the 0--10 s analysis window.} Reconstructed latencies outside this window were excluded. The mean across all 3,838 reconstructed arrivals was 4.781 s. The confidence interval uses standard errors clustered by bargaining group.}
    \label{tab:timing}
    \small
    \begin{tabular}{rrrl}
        \hline
        \textbf{Reconstructed} & \textbf{Within 0--10 s} & \textbf{Excluded} & \textbf{Mean (s) [95\% CI]}\\
        \hline
        3838 & 3784 & 54 & 4.618 [4.469, 4.767]\\
        \hline
    \end{tabular}
\end{table}

The ten-second hold was enforced by a countdown running in each participant's browser rather than on the server, so reconstructed latencies slightly above 10 s can arise from timer drift, and a browser that missed the end-of-round signal could continue to register moves after the round had ended. We traced the 54 excluded latencies against the server event log. Thirty correspond to a true pair-to-third gap of at most 10.5 s: 17 arrivals coincided with the expiry of the countdown, and in 13 rounds the pair had dissolved through an undo and re-formed, which the move-level reconstruction does not register. Seven are moves logged after the round had already ended, five are unbroken pairs whose countdown fired late (11.4 to 16.7 s), and 12 could not be verified because their event log was malformed. The results are insensitive to the rule: retaining all 54 at their reconstructed latencies gives a mean of 4.781 s (95\% CI [4.608, 4.954]).

\subsubsection*{Additional analyses of proposal success and coalition instability}

\begin{figure}
	\centering
	\includegraphics[width=0.8\textwidth]{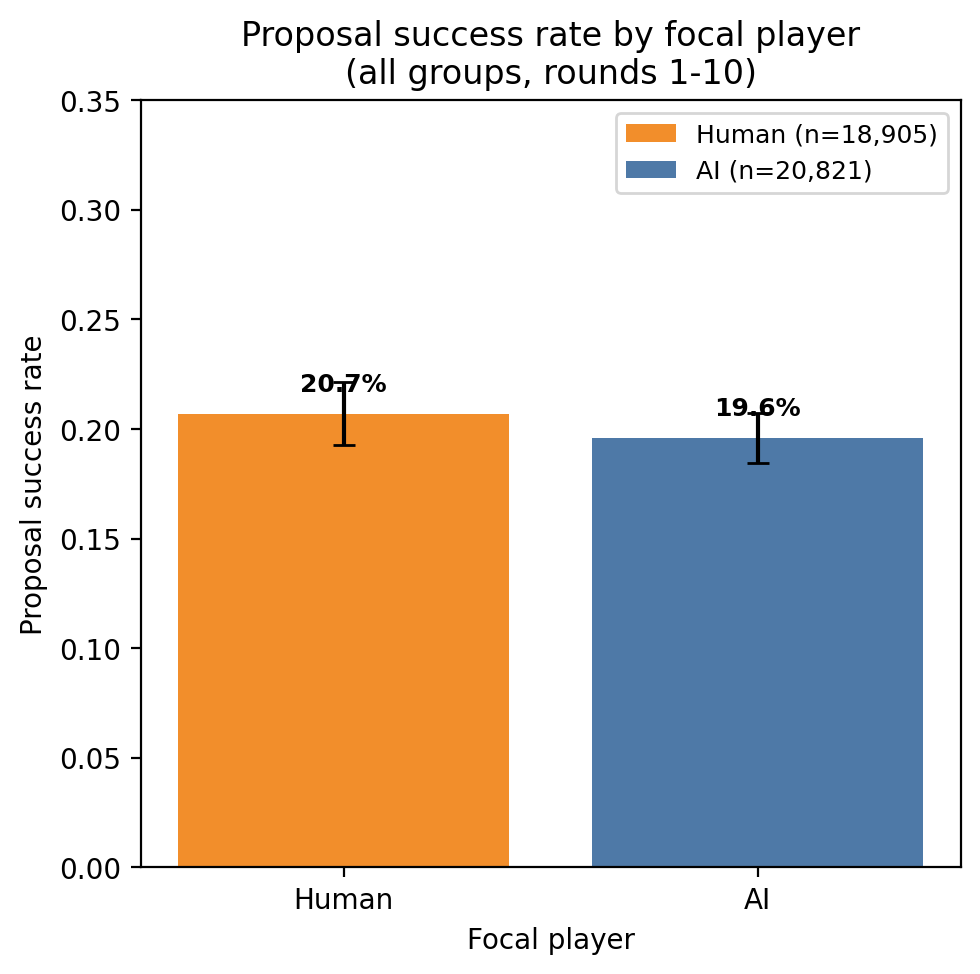}
	\caption{\textbf{Pooled proposal-success rate by focal player label.} Proposal success is the proportion of proposals that subsequently produced an agreement. Legend counts denote proposals. Error bars show 95\% confidence intervals from standard errors clustered by bargaining group.}
	\label{fig:ssuccess}
\end{figure}

Proposal success was defined as the proportion of proposals that subsequently produced an agreement. Pooled success was 20.7\% for Human players (3,913 of 18,905 proposals) and 19.6\% for AI players (4,081 of 20,821). Composition-specific differences were heterogeneous: in AAH groups, success was 16.4\% for Human and 21.3\% for AI focal players; in AHH groups, it was 22.6\% and 20.0\%, respectively. These comparisons are exploratory and do not identify whether label, offer content, or coalition opportunity caused success (Fig.~\ref{fig:ssuccess}).

\begin{figure}
	\centering
	\includegraphics[width=1.0\textwidth]{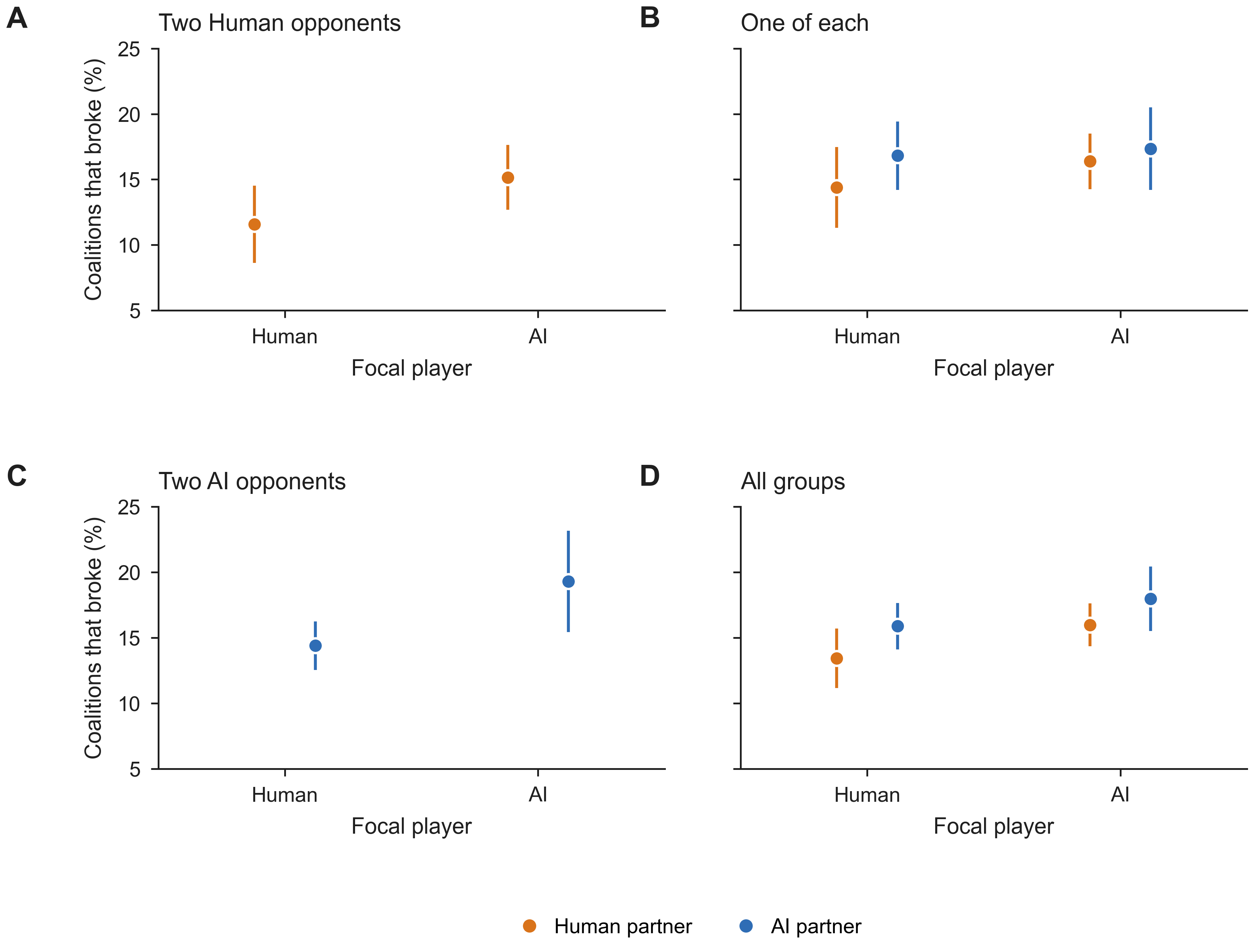}
	\caption{\textbf{Percentage of observed coalitions that subsequently broke, by focal player and partner label.} (\textbf{A} to \textbf{C}) By the composition of the focal player's two opponents; (\textbf{D}) all groups. Error bars show 95\% confidence intervals from standard errors clustered by bargaining group.}
	\label{fig:sbroken}
\end{figure}

Broken-coalition rates were 13.4\% when both the focal player and partner were Human, 15.9\% for a Human focal player with an AI partner, 16.0\% for an AI focal player with a Human partner, and 18.0\% when both were AI. We preregistered coalition instability as an exploratory analysis (Fig.~\ref{fig:sbroken}).

\subsubsection*{Post-game survey}

\begin{figure}
	\centering
	\includegraphics[width=1.0\textwidth]{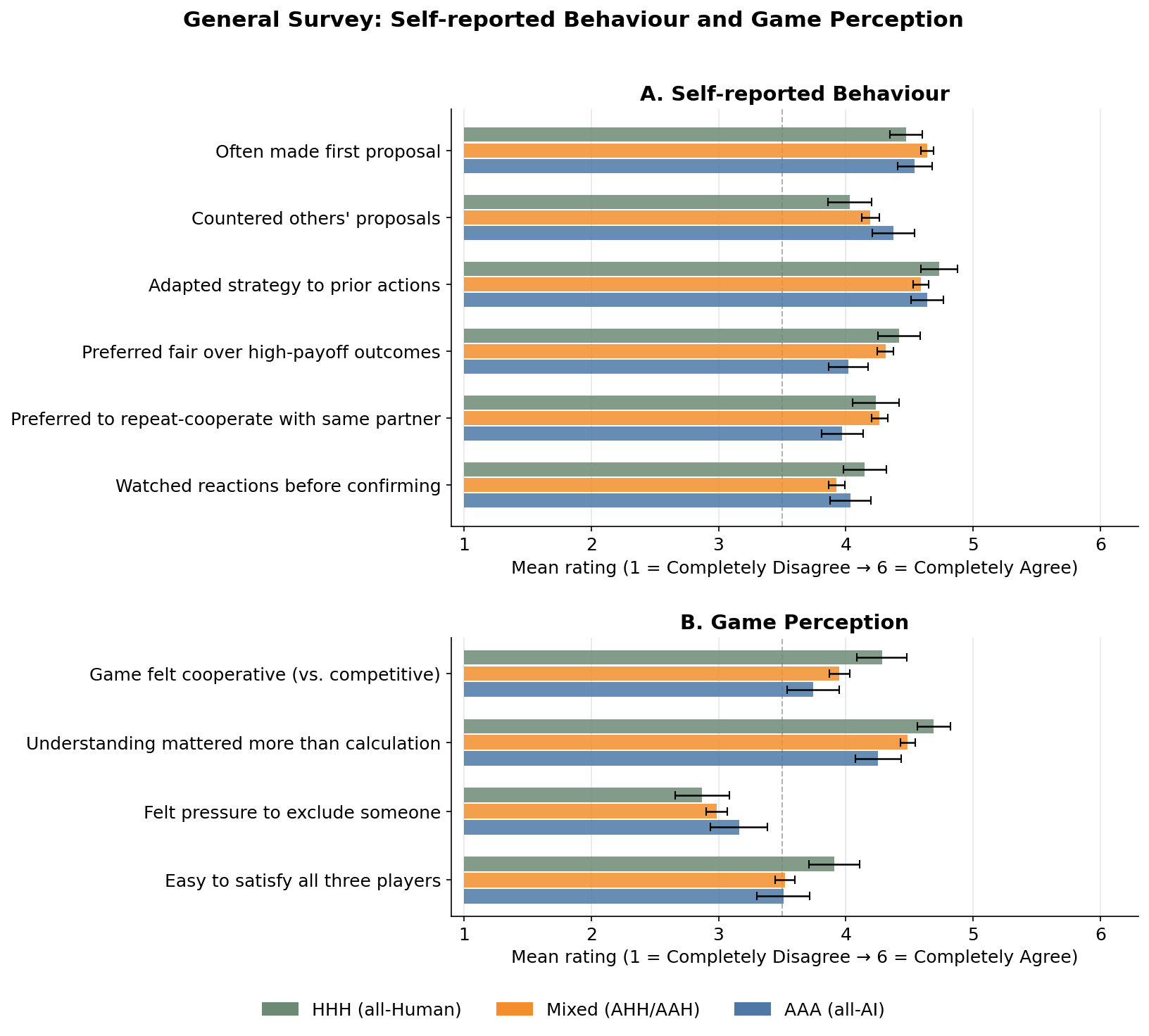}
	\caption{\textbf{Self-reported behaviour and game perception by group composition.} Error bars show 95\% confidence intervals from standard errors clustered by bargaining group.}
	\label{fig:ssurvey}
\end{figure}

Survey responses were available for about 2,225 participants, with AI-specific items administered to participants in mixed or all-AI conditions. Participants reported high awareness of the displayed labels (mean = 5.52 on the six-point scale), moderate willingness to delegate future negotiations to AI (mean = 4.11), and moderate self-reported strategy change based on label (mean = 3.81). Summary responses are reported in Table~\ref{tab:survey} and Fig.~\ref{fig:ssurvey}. These self-reports are descriptive and were not used to determine inclusion beyond the manipulation check.

\begin{table}
	\centering
	\caption{\textbf{Selected post-game survey responses.} Items used six-point agreement scales. AI-specific items were administered only to participants in mixed or all-AI conditions.}
	\label{tab:survey}
	\small
	\begin{tabular}{lrrr}
		\\
		\hline
		\textbf{Survey item} & \textbf{Mean} & \textbf{SD} & \textbf{N}\\
		\hline
		Often made first proposal & 4.61 & 1.21 & 2,227\\
		Countered others' proposals & 4.19 & 1.28 & 2,226\\
		Adapted strategy to prior actions & 4.61 & 1.20 & 2,225\\
		Preferred fair over high-payoff outcomes & 4.29 & 1.34 & 2,225\\
		Preferred repeat cooperation & 4.23 & 1.36 & 2,225\\
		Watched reactions before confirming & 3.97 & 1.35 & 2,225\\
		Aware of AI/Human labels & 5.52 & 0.95 & 1,688\\
		Willing to delegate future negotiations to AI & 4.11 & 1.59 & 1,688\\
		Changed strategy based on label & 3.81 & 1.69 & 1,686\\
		\hline
	\end{tabular}
\end{table}

\subsubsection*{Robustness to participant demographics}
As exploratory analyses that were not preregistered, we examined whether the results for Hypotheses 1 to 5 depend on the demographic characteristics of the focal participant recorded in the Prolific profiles (Table~\ref{tab:demographics}). First, we added sex, age group, ethnicity, student status, and employment status as covariates to the model reported for each hypothesis. The key coefficients changed by less than 0.01 (Table~\ref{tab:demorobust}). Second, we tested whether the focal effect for each hypothesis varied with each demographic characteristic, considering one characteristic at a time. Of the 25 interaction tests, five had $p < 0.05$, and one was below the Bonferroni threshold for 25 tests ($p < 0.002$): the premium for Human opponents in first offers (Hypothesis 3) was smaller among older proposers ($b = -0.144$ points per ten years of age, 95\% CI [$-0.232$, $-0.057$], $p = 0.001$). The premium was nonetheless present in every age group: it was 0.859 points (95\% CI [0.626, 1.092]) among proposers younger than 30 and 0.467 points (95\% CI [0.216, 0.718]) among those aged 50 or older. The other nominally significant interactions (age for Hypothesis 1, ethnicity and student status for Hypothesis 3, and ethnicity for Hypothesis 4) did not survive correction, and we do not interpret them.

\begin{table}
	\centering
	\caption{\textbf{Demographic composition of the analytical sample.} Counts with percentages of all participants in the column in parentheses, unless stated otherwise. Demographic information comes from the participants' Prolific profiles. Human and AI refer to the label under which a participant was presented to the other group members. ``Not available'' covers values that Prolific no longer returned and participants without a matching Prolific record.}
	\label{tab:demographics}
	\small
	\begin{tabular}{lrrr}
		\\
		\hline
		& \textbf{All} & \textbf{Human} & \textbf{AI}\\
		\hline
		Participants & 2,234 & 1,116 & 1,118\\
		Age in years, mean (SD) & 38.3 (12.5) & 38.3 (12.5) & 38.4 (12.6)\\
		Age in years, median (range) & 37 (18 to 88) & 36 (18 to 81) & 37 (18 to 88)\\
		\hline
		\textbf{Age group} & & &\\
		\quad 18--24 & 295 (13.2) & 156 (14.0) & 139 (12.4)\\
		\quad 25--34 & 675 (30.2) & 323 (28.9) & 352 (31.5)\\
		\quad 35--44 & 602 (26.9) & 305 (27.3) & 297 (26.6)\\
		\quad 45--54 & 391 (17.5) & 191 (17.1) & 200 (17.9)\\
		\quad 55--64 & 184 (8.2) & 98 (8.8) & 86 (7.7)\\
		\quad 65 or older & 76 (3.4) & 36 (3.2) & 40 (3.6)\\
		\quad Not available & 11 (0.5) & 7 (0.6) & 4 (0.4)\\
		\hline
		\textbf{Sex} & & &\\
		\quad Female & 1,086 (48.6) & 540 (48.4) & 546 (48.8)\\
		\quad Male & 1,118 (50.0) & 560 (50.2) & 558 (49.9)\\
		\quad Prefer not to say & 14 (0.6) & 6 (0.5) & 8 (0.7)\\
		\quad Not available & 16 (0.7) & 10 (0.9) & 6 (0.5)\\
		\hline
		\textbf{Ethnicity} & & &\\
		\quad White & 1,497 (67.0) & 754 (67.6) & 743 (66.5)\\
		\quad Black & 241 (10.8) & 114 (10.2) & 127 (11.4)\\
		\quad Asian & 237 (10.6) & 113 (10.1) & 124 (11.1)\\
		\quad Mixed & 143 (6.4) & 76 (6.8) & 67 (6.0)\\
		\quad Other & 68 (3.0) & 32 (2.9) & 36 (3.2)\\
		\quad Prefer not to say & 21 (0.9) & 15 (1.3) & 6 (0.5)\\
		\quad Not available & 27 (1.2) & 12 (1.1) & 15 (1.3)\\
		\hline
		\textbf{Student} & & &\\
		\quad Yes & 329 (14.7) & 162 (14.5) & 167 (14.9)\\
		\quad No & 1,587 (71.0) & 789 (70.7) & 798 (71.4)\\
		\quad Not available & 318 (14.2) & 165 (14.8) & 153 (13.7)\\
		\hline
		\textbf{Employment} & & &\\
		\quad Full-Time & 948 (42.4) & 464 (41.6) & 484 (43.3)\\
		\quad Part-Time & 337 (15.1) & 183 (16.4) & 154 (13.8)\\
		\quad Unemployed (and job seeking) & 321 (14.4) & 153 (13.7) & 168 (15.0)\\
		\quad Not in paid work (e.g. homemaker, retired or disabled) & 183 (8.2) & 91 (8.2) & 92 (8.2)\\
		\quad Other & 102 (4.6) & 54 (4.8) & 48 (4.3)\\
		\quad Due to start a new job within the next month & 28 (1.3) & 10 (0.9) & 18 (1.6)\\
		\quad Not available & 315 (14.1) & 161 (14.4) & 154 (13.8)\\
		\hline
	\end{tabular}
\end{table}

\begin{table}
	\centering
	\caption{\textbf{Robustness of Hypotheses 1 to 5 to participant demographics.} Key coefficient of the model reported for each hypothesis, with 95\% confidence intervals, as reported and after adding the focal participant's sex, age group, ethnicity, student status, and employment status as covariates (missing values as a separate category). The last five columns give p-values for tests of whether the focal effect varies with each demographic characteristic, each from a separate model fitted to participants for whom that characteristic is recorded (joint Wald test for characteristics with more than two categories; age in years as a linear term; ``prefer not to say'' treated as not recorded; employment in five categories, with those about to start a job counted under Other). Key coefficients: H1, difference in payoff between the Human and the AI label; H2, difference in the log-odds of a Human partner between Human and AI focal players; H3, difference in first-offer points between Human and AI opponents; H4, difference in retained points between Human and AI partners; H5, interaction of the Human label with the Delegation framing on payoff. Standard errors are clustered by bargaining group. Exploratory; p-values are not corrected for the 25 tests.}
	\label{tab:demorobust}
	\footnotesize
	\setlength{\tabcolsep}{3.5pt}
	\begin{tabular}{lllrrrrr}
		\\
		\hline
		& & & \multicolumn{5}{c}{\textbf{Interaction, p}}\\
		\textbf{Hypothesis} & \textbf{Reported} & \textbf{Adjusted} & \textbf{Sex} & \textbf{Age} & \textbf{Ethn.} & \textbf{Student} & \textbf{Empl.}\\
		\hline
		H1: payoff gap & 0.150 [0.112, 0.188] & 0.149 [0.111, 0.187] & 0.512 & 0.023 & 0.552 & 0.231 & 0.419\\
		H2: partner choice & 0.424 [0.151, 0.697] & 0.429 [0.156, 0.703] & 0.219 & 0.364 & 0.668 & 0.136 & 0.083\\
		H3: first offers & 0.728 [0.609, 0.847] & 0.727 [0.608, 0.846] & 0.131 & 0.001 & 0.030 & 0.049 & 0.101\\
		H4: retained points & $-$0.179 [$-$0.386, 0.027] & $-$0.182 [$-$0.389, 0.025] & 0.268 & 0.736 & 0.036 & 0.709 & 0.752\\
		H5: framing & $-$0.007 [$-$0.084, 0.069] & 0.000 [$-$0.075, 0.076] & 0.687 & 0.085 & 0.657 & 0.417 & 0.671\\
		\hline
	\end{tabular}
\end{table}
\end{document}